\documentclass[journal=jacsat,manuscript=article]{achemso}
\usepackage[utf8]{inputenc}
\usepackage[T1]{fontenc}
\usepackage[usenames,dvipsnames]{xcolor}
\usepackage[english]{babel}
\usepackage{graphicx}
\usepackage{rotating} %sidewaystable
\usepackage{float}
\usepackage{placeins}
\usepackage[colorlinks, linkcolor = black, citecolor = black, filecolor = black,
urlcolor =black]{hyperref}
\usepackage{subfigure}
\usepackage{braket}
\usepackage{amsmath,mathtools}

\title{Time-Dependent Atomistic Simulations of the  CP29 Light-Harvesting Complex}

\author{Sayan Maity}
\affiliation{Department of Physics and Earth Sciences, Jacobs University Bremen, Campus Ring 1, 28759 Bremen, Germany}

\author{Pooja Sarngadharan}
\affiliation{Department of Physics and Earth Sciences, Jacobs University Bremen, Campus Ring 1, 28759 Bremen, Germany}

\author{Vangelis Daskalakis}
\affiliation{Department of Chemical Engineering, Cyprus University of Technology, 30 Archbishop Kyprianou Str. 3603, Limassol, Cyprus}

\author{Ulrich Kleinekath\"ofer}
\affiliation{Department of Physics and Earth Sciences, Jacobs University Bremen, Campus Ring 1, 28759 Bremen, Germany}
\email{u.kleinekathoefer@jacobs-university.de}

\begin{document}
	
\begin{abstract}
Light harvesting as the first step in photosynthesis is of prime importance for
life on earth. For a theoretical description of  photochemical processes during
light harvesting, spectral densities are key quantities. They serve as input
functions for modeling the excitation energy transfer dynamics and spectroscopic
properties. Herein, a recently developed  procedure is applied to determine the
spectral densities of the pigments in the minor antenna complex CP29  of
photosystem II which has recently gained attention because of its active role in
non-photochemical quenching processes in higher plants. To this end, the density
functional based tight binding (DFTB) method has been employed to enable
simulation of the ground state dynamics in a quantum-mechanics/molecular
mechanics (QM/MM) scheme for each chlorophyll pigment. Subsequently, the
time-dependent extension of the long-range corrected DFTB approach has been used
to obtain the excitation energy fluctuations along the ground-state trajectories
also in a QM/MM setting. From these results the spectral densities have been
determined and compared for different force fields and to spectral densities
from  other light-harvesting complexes. In addition, the excitation energy
transfer in the CP29 complex has been studied using ensemble-average wave packet
dynamics.
\end{abstract}

\section{Introduction} Chlorophylls (Chl), bacterio-chlorophylls (BChl) and
carotenoids (Car) are the key pigment molecules involved in excitation energy
transfer (EET) processes in light-harvesting (LH) complexes of plants and
bacteria. Actively involved are mainly the low-lying excited states of these
pigment molecules.  Aim of these processes is to transport the energy absorbed
from the sunlight to reaction centers where charge separation takes place as one
of the steps of photosynthesis \cite{blan14a}. In the last two decades, the
interest in some LH complexes of bacteria and marine algae has been spurred due
to the claim of experimentally observed  long-lived quantum coherences in EET
processes at low as well as at ambient
temperatures\cite{enge07a,coll09a,coll10a,pani10a,cao20a}. Recently, however it
became clear that these long-lived oscillations likely originate from
impulsively excited vibrations and are too short-lived to have any functional
significance in photosynthetic energy transfer \cite{duan17b,thyr18a,cao20a}.

Independent of these developments, the attention towards plant LH systems has
significantly increased in recent years. Especially the topic of photoprotection
became a field of growing interest. Under the stress of excess solar energy, LH
complexes of plants activate a mechanism termed non-photochemical quenching
(NPQ) which can be invoked to avoid
photo-inhibition\cite{ruba07a,ruba12a,chme16a}. Excess sunlight leads to an
enlarged pH gradient across the thylakoid membrane containing the LH complexes
of the photosystem II (PSII). An increased pH gradient triggers the switching
between the photochemical light harvesting and a non-photochemical quenching
mode\cite{tian19a,nico19a,buck19a,dela19a}. Moreover, not only the pH gradient
but also the presence of a protein termed photosystem II subunit S (PsbS)
induces conformational changes in the antenna complexes and thus take part in
activating the quenching mechanism\cite{li00a,corr16a,ligu19a,guar20a}. The
detailed changes in the conformations are, however,  still not well understood
at the molecular level. The excess sunlight absorbed by the Chl molecules gets
released as heat via carotenoid pigments in a mechanism that
is a topic of active research \cite{ruba07a,tian19a,ruba18a}. Based on experimental findings, it  has been claimed that the major antenna LHCII and the minor antennas, in
particular the CP29 complex,  play a crucial role in protecting the
photosynthetic apparatus of PSII from  excess excitation
energy\cite{dall17a,nico19a,guar20a,tian19a,chme16a,son19a}.

In order to study the dynamics of the LH systems of plants, various models were
built based on  crystal structures of the
LHCII\cite{liu04a,muh10a,muh12a,duff13a,chme15a} and CP29
complexes\cite{pan11a,wei16a,mueh14a,juri15b,fox18a,lapi20a}. Furthermore,
additional models were created taking into account the lumenal pH gradient and
the presence of PsbS proteins trying to realistically mimic the NPQ
process\cite{dask19b,dask19a,mait19a,dask18a,dask20a}. In these models, either
the population transfer of excitons\cite{krei14a,rode16a} or the associated
rates were determined\cite{dask19b,lapi20a}. The major input parameters which
are required in such calculations are the excitation energies of the individual
pigments, also known as site energies, the excitonic couplings between the
pigments and the so-called spectral densities. In the present study, the latter
quantity  is in the focus while also some of the parameters are being
determined.
\begin{figure} [tb]
\centering \includegraphics[width=0.85\textwidth]{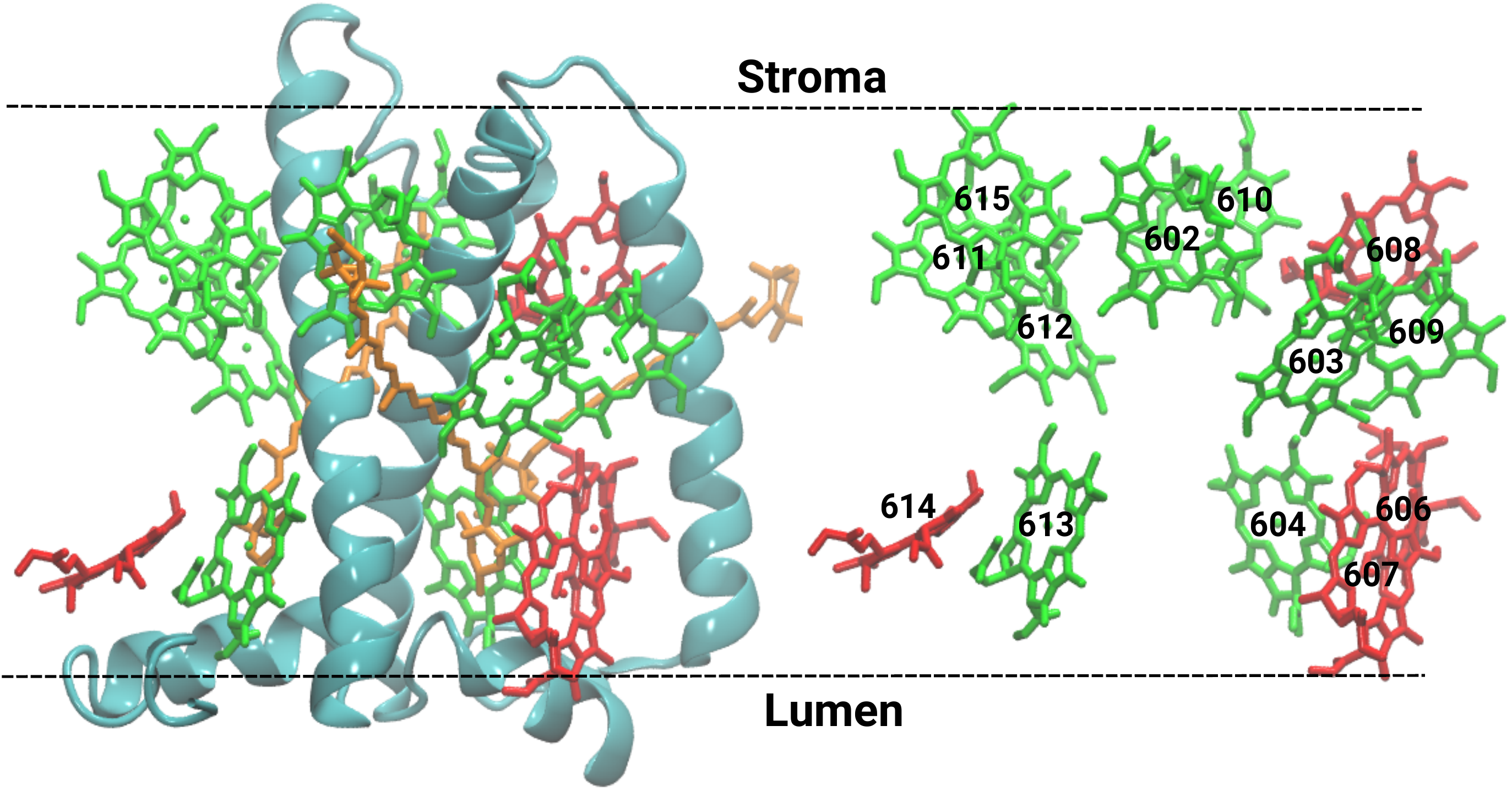}
\caption{\label{fig:sys} The protein part of the  CP29 minor antenna complex of
	spinach represented in an iceblue cartoon representation. The Chl-a and Chl-b
	molecules are depicted in green and red while the carotenoids Lut, Vio and Neo
	are shown in orange. In addition, the arrangement of the chlorophylls is
	displayed in the right panel together with the respective residue numbering
	according to the crystal structure\cite{pan11a} (pdb code: 3PL9).}
\end{figure}	

The spectral density is the key ingredient for performing exciton dynamics of LH
complexes within the framework of open quantum systems\cite{may11}. It accounts
for effects of the pigment environment but also  internal modes on the energy
gap between ground and first excited states. The accurate determination of
spectral densities is, however, a challenging task especially for LH systems due
to the complexity of these entities and the numerical effort involved. One
approach to determine spectral densities which has been employed in several
previous investigations\cite{damj02a,olbr10a,olbr11b,shim11a,aght14a} is to
determine the excitation energy gap along a classical MD trajectory using
ZINDO/S-CIS (Zerner's Intermediate Neglect of Differential Orbital method with
spectroscopic parameters together with the configuration interaction using
single excitation) or TDDFT (time-dependent density functional theory)
calculations. This approach, however, has problems especially in the high
frequency region of the respective spectral densities \cite{lee16a,mait20a}. The
shortcoming is due the inability of classical force fields to accurately
described the vibrational modes of the pigments as well as to provide a proper
sampling of the  geometrical phase space. Inaccuracies in the ground state
conformations and dynamics are subsequently passed on to the determination of
the energy gap fluctuations. The inconsistency in geometry between the ground
and excited state is commonly known as ``geometry mismatch'' problem
\cite{curu17a}. Some recent methods to determine the intramolecular vibrational
modes  accurately are based on normal mode analyses \cite{lee16a,lee16c} or
ground state dynamics on pre-calculated quantum mechanical potential energy
surfaces\cite{kim16a,kim18a} and have been employed in calculations of spectral
densities. In spite of providing impressive agreements compared to experimental
results, these aforementioned schemes are computationally still demanding
\cite{kim16a,padu17a}. A sophisticated quantum-mechanics/molecular mechanics
(QM/MM) MD with an accurate description of the vibrational properties of the
pigments would be an alternative way, however, semi-empirical schemes have a
limited accuracy\cite{rosn15a} while DFT-based calculations are numerically
 expensive for pigments in LH systems\cite{blau18a}. To this end, we
recently proposed a scheme  using   QM/MM MD dynamics employing the numerically
efficient density functional based tight binding (DFTB) approach\cite{elst98a}
and have shown to be able to obtain a  good agreement between the  spectral
densities obtained in such a manner and their experimental
counterparts\cite{mait20a,mait21a}.

In this study, we have applied a recently developed multiscale protocol to
determine the spectral densities of the individual pigments within the minor
antenna complex CP29 of the PSII system. The CP29 complex contains a total of
thirteen chlorophyll molecules including nine Chl-a and four Chl-b. In addition,
three different types of carotenoids, i.e.,  lutein (Lut), violaxanthin (Vio)
and neoxanthin (Neo) are present in this antenna complex (see
Fig~\ref{fig:sys}). The Chl-b chromophores in the  periphery of the protein
matrix act as accessory pigments transferring the absorbed light energy towards
the pool of Chl-a molecules quite rapidly since the excited states of the latter
pigments are slightly lower in energy. From the Chl-a molecules, the energy
obtained during light harvesting is moved on to the PSII reaction center. While
the carotenoids are speculated to participate in the quenching process, here the
aim is to  accurately determine the spectral densities for the key pigments,
i.e., the chlorophyll molecules in the  CP29 complex. To this end, the AMBER
force field has been employed in connection with the DFTB approach in a QM/MM
framework. Details of the computational scheme can be found in the following
method section. The spectral density profiles obtained in the present
calculations show a very good agreement with the experimental counterparts in
line with our previous work for other LH systems \cite{mait20a,mait21a}.
Furthermore, the impact of the choice of the classical force field in the QM/MM
setting has been investigated by redoing the calculations using the OPLS force
field in connection with the DFTB method. Only very small differences can be
seen in the spectral density profiles when using the two different force fields
within the QM/MM framework. This finding indicates that both force fields  are
suitable for the modeling LH complexes. The present study also incorporates a
brief comparison of the spectral densities of the BChl and Chl pigments in
different LH systems. Before concluding, we report on time-averaged and
time-dependent exciton Hamiltonians and on ensemble-averaged wave-packet
dynamics within the Ehrenfest formalism for the exciton transfer in the CP29
complex.

\section{Computational Method} 
\begin{figure} [tb]
\centering \includegraphics[width=0.5\textwidth]{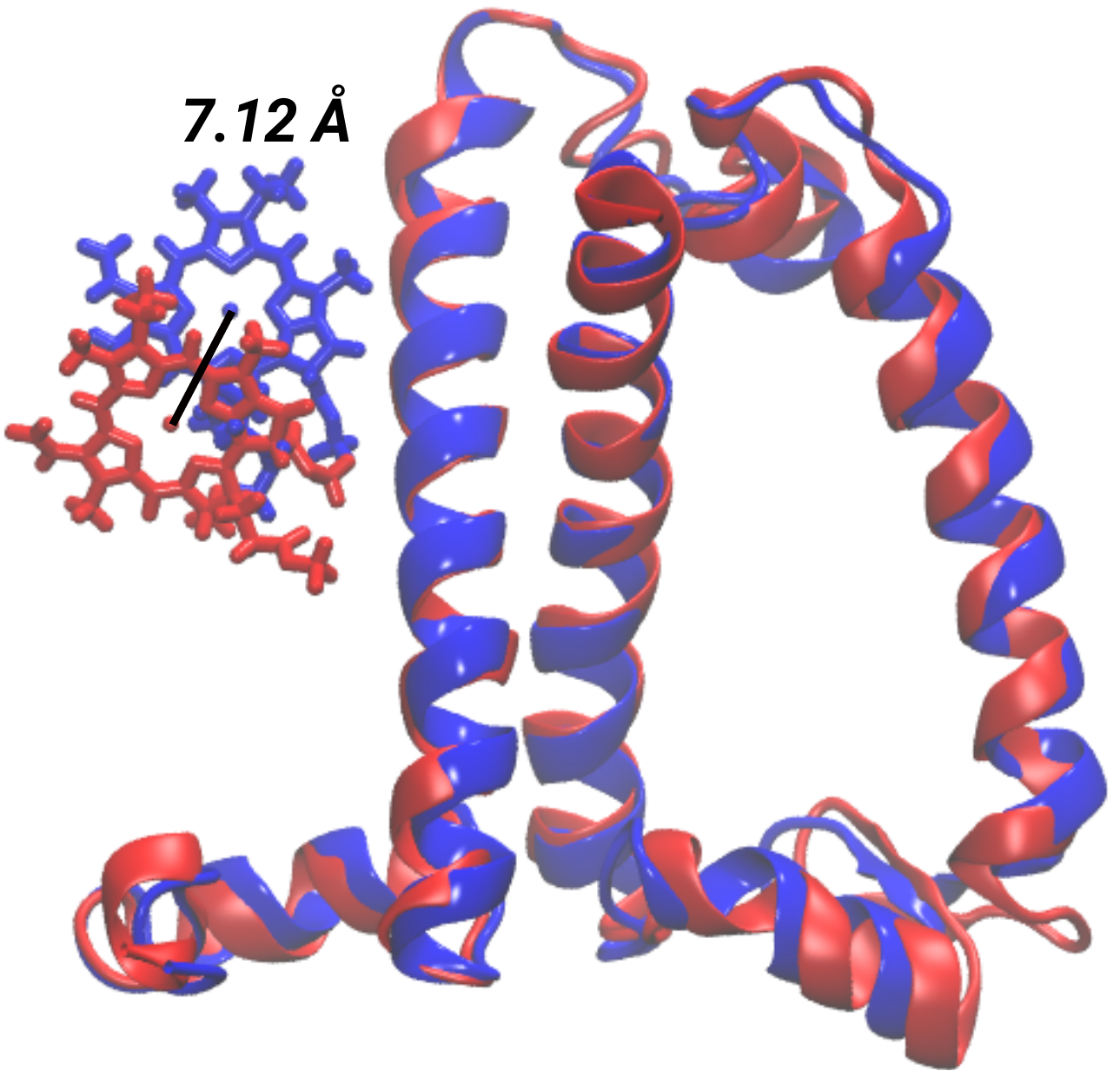}
\caption{\label{fig:shift} Structural shift of the pigment Chl-a 615 after the
	300~ns classical unbiased simulation (red) compared to the crystal structure
	used as starting structure (blue).}
\end{figure}

The X-ray crystal structure of the CP29 antenna
complex from spinach (pdb code: 3PL9)\cite{pan11a} has been used  as the
starting point for the atomistic modeling. The first 87 residues were not
resolved in the X-ray structure and  neglected in the present study since their
influence on the spectral densities is assumed to be small. Furthermore, the
glyceraldehyde 3-phosphate (G3P) molecule which was tentatively found in-between
Chl-a 611 and 615 was not taken into account during the system preparation. The
AMBER03 force field\cite{duan03a} has been employed for the  protein using the
GROMACS-5.1.4 suite of programs\cite{abra15a} for the simulations. Moreover, the force field parameters for Chl-a/b \cite{cecc03a,zhan12a} and carotenoids\cite{pran16a} have been taken from previous studies. First, a POPC lipid membrane was prepared using the
CHARMM-GUI server \cite{jo08a}, then the topology for the membrane was generated
using the LEaP program of AmberTools-20 together with the Lipid-17 force field
of AMBER. Thereafter, the topology and coordinates  were transformed to the
GROMACS format with the  help of the ACEPYPE interface \cite{dasi12a}. The
simulation box size was chosen to be 115 $\times$ 115 $\times$ 90 {\AA} and the
system was solvated with TIP3P water molecules. Furthermore, four potassium ions
were added to neutralize the whole system. After preparing the system, an energy
minimization was performed using the steepest-descent algorithm to remove close
contacts. Subsequently,  a 2~ns NVT equilibration was carried out at 300 K with
a 1~fs integration time step, keeping  position restraints on protein, lipid and
cofactor atoms. Subsequently, a 20~ns NPT run was performed using the same time
step and position restraints. After that, another 10~ns NPT simulation was run
keeping the restraints on protein, pigments and the phosphorous atoms of the
lipid molecules. An additional 10~ns NPT run was carried out using a 1~fs time
step in which the first 5~ns were performed with restraints on protein and
pigments whereas for the  last 5~ns the restraints were only on the protein and
Chl molecules. Subsequently, in the next 5~ns NPT equilibration, the position
restraint was kept only on the protein and the Mg atoms of Chl pigments.
Subsequently, yet another 5~ns NPT run was performed using a 2~fs time step in
which the position restraints were kept on the protein for the first 2~ns, then
for 1~ns on the side chains atoms and  1~ns on the C$\alpha$ atoms of the
protein. Finally, a 50 ns long NPT unbiased simulation was carried out without
any position restraints and the coordinates were stored using a 25~ns stride
which were subsequently employed as starting structures for the QM/MM MD
simulations in the next step. The LINCS algorithm was applied and the
constraints were kept on the H-bonds throughout the equilibration process. The
equilibrated structure was further equilibrated for another 200~ns  where the
coordinates were stored at 20~ps. This produces 10,000 frames which were
utilized for the excitonic coupling calculations using the TrESP (transition
charges from electrostatic potentials) approach \cite{madj06a,olbr11a} in order
to construct the system Hamiltonian. A shift in the position  of pigment Chl-a
615 was observed after the classical simulations (see Fig.~\ref{fig:shift}).
Structurally, this pigment is  solvent exposed in case of an isolated CP29
crystal structure and thus can more easily move than a pigment embedded in the
middle of the protein. A movement by 3.97 {\AA}  was found after the 100 ns
equilibration whereas after the 200 ns production run the movement was by 7.12
{\AA}. This finding is consistent with the latest cryo-EM structure of the
C$_2$S$_2$M$_2$ supercomplex \cite{su17a}, in which a different Chl-a 601 was
found close to the position where Chl-a 615 was found in the X-ray structure
used in the present study \cite{pan11a}. Thus, we assume that the contribution
of Chl-a 615 will be similar to that of the Chl-a 601 found in the latter case.

Two sets of QM/MM MD simulations, based on starting geometries extracted at 25
and 50~ns from the unbiased classical simulations, were performed to minimize
the sampling problems of spectral densities found earlier for the present kind
of simulations\cite{mait20a}. A 80~ps NPT QM/MM dynamics was performed with a
0.5~fs integration time step without any bond constraints for the Chl-a and
Chl-b molecules. The phytyl tail was truncated at the so-called ``C1-C2'' bond
and capped by a hydrogen link atom. This truncation is sensible since the tails
play no role in the $\pi$-conjugated rings but their partial charges
potentially influence the $Q_y$ excitation energies of the Chl molecules and
thus need to be included in the MM part.  Such a  treatment significantly reduces the
computational costs compared to a full QM treatment of the Chl
molecules\cite{bold20a}. The DFTB3 theory\cite{gaus11a} with the frequency
corrected 3OB parameter set (3OB-f)\cite{gaus13a} was employed coupled to an
AMBER force field as implemented in the GROMACS-DFTB+ interface for the ground
state dynamics\cite{kuba15b,hour20a}. The 3OB-f parameter set was  specially
designed in connection with the so-called 3OB parameters to describe the 
vibrational frequencies of C=C, C=N and C=O bond stretching modes  more
accurately. These modes are very relevant for molecules including  porphyrins
ring such as Chl pigments. The last 60~ps of each trajectory were  stored with
a stride of 1~fs. This procedure results in 2 $\times$ 60,000 frames for
each Chl molecule using the two different starting structures of the QM/MM MD
trajectories. In addition, we extended the first set of the simulations up to
1.1 ns with a 1 fs integration time step and stored the geometries for the last
1 ns dynamics with a stride of 100~fs.  Thus, another 10,000 frames were
generated which were mainly utilized to calculate the average site energies of
the CP29 complex. Furthermore,  the CP29 complex, as modeled using the OPLS force
field in a previous study\cite{dask18a}, was  used as
starting structure for QM/MM MD simulations. Again, two different starting
structures were considered as the initial geometries for the QM/MM dynamics.
This time the starting structure were 10~ns apart in the unbiased
trajectories. Again we have stored 2 $\times$ 60,000 frames for each Chl
pigment in order to perform the spectral density calculations.

The snapshots collected from the DFTB-QM/MM MD trajectories were utilized to
calculate the $Q_y$ excitation energies employing the time-dependent extension
of the long-range corrected DFTB (TD-LC-DFTB) in a QM/MM fashion. The LC-DFTB method
is the DFTB analogue of the long-range corrected DFT (LC-DFT) approach with
functionals like CAM-B3LYP and $\omega$B97X  which has lately been developed
to overcome problems related to charge transfer and overpolarization in
conjugated systems\cite{kran17a}. In recent studies, we have shown that
TD-LC-DFTB is a close to ideal alternative to DFT with long-range corrected functionals
in order to compute excitation energies and excitonic couplings in a numerically
efficient way\cite{bold20a}. To this end, we have employed the DFTB+ package in
which the TD-LC-DFTB scheme is implemented  based on the OB2 parameter
set\cite{kran17a,hour20a}. Moreover, during the excitation energy calculations
the QM region is shifted towards the center of the simulation box to avoid 
boundary problems during the non-periodic QM/MM calculations. After capturing the
excitation energies along the QM/MM trajectories, these were used to determine
the  autocorrelation functions from which the spectral densities were calculated
by performing Fourier transformations as described in below.

The excitonic couplings have been determined based on the 200~ns classical MD
simulation. 10,000 snapshots were collected and utilized for the TrESP excitonic
coupling calculations as described below. For this purpose, the transition
charges  for the Chl-a and Chl-b molecules were determined with the help of the
Multiwfn\cite{lu12a} and the ORCA\cite{nees18a} packages. Transition charges of
these molecules exist \cite{mueh14a}, but we repeat them here with a long-range
corrected functional and a larger basis set. First, a geometry optimization was
performed at the B3LYP level of theory together with a def2-TZVP basis set as
implemented in the ORCA program. Furthermore, the resolution of identity RIJCOSX
was employed together with the auxiliary basis set def2/J in order to speed up
the calculations. In a subsequent step, the optimized geometries were utilized
to performed excited state TD-DFT calculation employing the Tamm-Dancoff
approximation at the CAM-B3LYP level of theory as implemented in ORCA. The same
basis set and the resolution identity were applied again for these calculation.
Finally, the transition densities obtained from the TD-DFT calculations were
used in the Multiwfn package for the electrostatic fitting. During the fitting
procedure, the charges on the hydrogen atoms were set to zero and the transition
densities were distributed among the heavier atoms of the Chl-a and Chl-b
molecules. The obtained TrESP charges are listed in the  SI (see Table~S1) and were used in
the TrESP calculations for the excitonic couplings. Moreover, we have employed
a  scaling factor of 0.81 for the Chl-a charges and of 0.83 for the Chl-b
charges in order to reproduce the experimental transition dipole moments of 5.7
D and 4.6 D, respectively\cite{knox03a}.

\subsection{Theoretical Background} To be able to determine the exciton dynamics
in LH complexes within a tight-binding model,  one needs to construct the
excitonic system Hamiltonian based on the site energies $E_m$ of the pigments
$m$ and the respective couplings $V_{nm}$  \cite{may11}
\begin{equation}
H_S = \sum_{m} E_m \ket{m} \bra{m} + \sum_{n \ne m} V_{mn} \ket{n} \bra{m}~.
\label{eq:Hamil}
\end{equation}
If the site energies and coupling are determined based on (quantum) molecular
dynamics trajectories, these quantities are time-dependent. Two
main options are available how to determine the exciton dynamics. One option is to perform
calculations directly based on these time-dependent Hamiltonians, e.\ g., some
kind of ensemble-averaged Ehrenfest approach (without back-reaction on the
bath)\cite{aght12a} sometimes also termed NISE (numerical integration of the
Schrödinger equation) \cite{jans18a}.  Alternatively, one can average the
elements of the Hamiltonian over time. The site energy fluctuations are then
 represented by the so-called spectral density. In principle, one can
also define spectral densities for the coupling fluctuations but this is very
rarely done since the effect of these fluctuations on the dynamics  is negligible\cite{aght17a}.
On the experimental side, spectral densities corresponding to those based on the
excitation energy fluctuations can be obtained using  delta fluorescence line
narrowing ($\Delta$FLN) spectroscopy. In the present study, we employ a cosine
transformation of the  energy autocorrelation functions decorated with a thermal
prefactor to determine the spectral density\cite{damj02a,olbr11b,vall12a}
\begin{equation}
J_m(\omega)=\frac{\beta\omega}{\pi} \int\limits_{0}^{\infty} dt~ C_m(t) \cos(\omega t)
\label{eq:spd}~.
\end{equation}
The necessary autocorrelation functions for each pigment $m$ can be written as
\begin{equation}
C_m(t_l) = \frac{1}{N-l} \sum_{k=1}^{N-l} \Delta E_m(t_l + t_k) \Delta E_m(t_k) ~.
\label{eq:acf}
\end{equation}
Here $\Delta E_m$ denotes the difference of  site energy $E_m$ from its average
value $E_m= E_m - \langle E_m \rangle$ and $N$  the number of snapshots present
in the respective part of the trajectory. Moreover, we follow the same procedure
as detailed in our previous work \cite{mait20a,mait21a} to obtain the final
correlation functions and spectral densities.

For computing excitonic couplings in LH complexes, the  TrESP method has been shown
to be accurate for medium and large distances.Once the atomic transition charges of the pigment molecules have been
determined, the coupling values  can be calculated as
\begin{equation}
V_{mn} = \frac{f}{4\pi\epsilon_0} \sum_{I,J}^{m,n} \frac{q_I^T \cdot q_J^T}{\mid r_m^I - r_n^J~, \mid}~.
\label{eq:tresp}
\end{equation}
where $q_{I}^T$ and $q_{J}^T$ denote the transition charges of atoms $I$ and $J$
and  $f$  a distance-dependent screen factor which taking environmental
influences on the excitonic coupling into account . Here, we have employed an
exponential screening factor derived by Scholes et al.~\cite{scho07a}
\begin{equation}
f(R_{mn}) = A \exp (-BR_{mn} + f_{0})~.
\label{eq:screen}
\end{equation} 
In this expression $A$, $B$, and $f_{0}$  have the values 2.68, 0.27 and 0.54,
respectively\cite{scho07a}.

\section{Results and Discussion} \subsection{Site Energy Calculations} 
\begin{figure} [tb]
\centering \includegraphics[width=\textwidth]{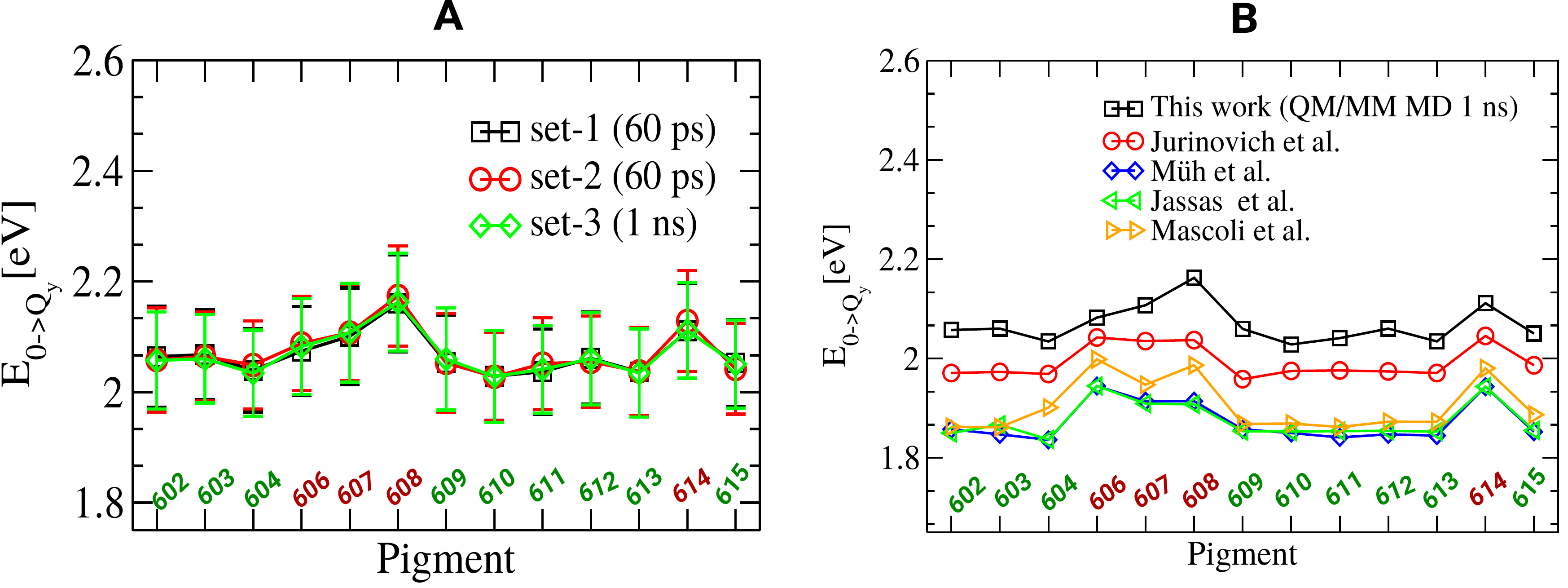}
\caption{\label{fig:energy} A) Average site energies of the Chl molecules in the CP29 complex with the respective error bars indicating the fluctuations along the three different QM/MM MD trajectories. The residues 602, 603, 604, 609, 610, 611,612, 613, and 615 refer to Chl-a molecules (green) while the residues 606, 607, 608, and 614 to are Chl-b molecules (red). B) Comparison with other computed  site energies for the CP29 complex by Jurinovich et al.\cite{juri15b}, Müh et al.\cite{mueh14a}, Jassas et al.\cite{jass18a} as well as Mascoli et al.\cite{masc20a}.}
\end{figure}

As
starting point of the analysis, the average site energies of the thirteen Chl
molecules have been determined for the three  QM/MM MD trajectories. The
structurally very similar Chl-a and Chl-b pigments contain  Mg-porphyrin rings
which are main responsible for the electronic properties of the respective
molecules and especially the excited $Q_y$ state. When the  methyl group of  a
Chl-a porphyrin ring is oxidized to become an aldehyde group, the molecules
becomes a Chl-b molecule with a  blue shift in the excitation energies (see Fig.~S1).
Due to the higher $Q_y$ excitation energies of Chl-b molecules, excitation energy from
these pigments will, depending on the respective couplings, flow to neighboring
Chl-a  molecules. The energy is subsequently shared between the Chl-a molecules
and transferred further into the direction of lower energies ending up at a
reaction center. In the present study, the TD-LC-DFTB approach has been employed
as the QM approach to compute the excitation energies for the individual Chl
molecules along the  QM/MM MD trajectories. As for basically all DFT approaches,
the energy gaps are overestimated which in line with our previous
observations\cite{bold20a,mait20a,mait21a}. In many cases, this does, however,
not cause problems since we are mainly interested in the relative site energies.
The energy ladder and the corresponding standard deviations due to the thermal
fluctuations along the two different 60 ps-long and the one 1 ns-long  QM/MM MD
trajectories are shown in Fig.~\ref{fig:energy}A.
The results for the average site energies of the individual pigments are almost
identical for the three different trajectories. Although, the 1~ns QM/MM MD
trajectory is averaging over a much longer time span than the 60~ps runs, no
significant differences are observed. The respective distributions for the site
energies which are also known as the densities of states (DOS) are depicted in
Figs.~S2 and S3. The shape of these distributions is Gaussian, i.e., symmetric, and not
skewed as observed in similar calculations using the semi-empirical ZINDO/S-CIS
approach\cite{olbr10a,aght14a}. As expected, the average site energies of the
Chl-b pigments are higher in energy than the Chl-a molecules although the
spectral densities for these two types of molecules are almost identical as
shown below. The structural difference between the Chl-a and b molecules has
already been mentioned above  and, as depicted in  Fig.~S1, is the only reason
behind the blue shifted site energies of Chl-b molecules. Furthermore, the
excitation energy fluctuations of the Chl-a molecules are in the same range as
those found for the same pigment type in the  LHCII complex\cite{mait21a}.
Moreover, we have compared the  calculated site energies with literature results
as shown in Fig.~\ref{fig:energy}B. In case of the results by the Jurinovich et
al.\cite{juri15b}, TD-DFT calculations were carried out along classical MD
trajectory based on the CAM-B3LYP/6-31G(d) level of theory within a polarizable
QM/MM description. Since CAM-B3LYP is a long-range corrected DFT functional, the
site energies obtained employing this approach, are quite similar to our
TD-LC-DFTB calculations. In case  of the calculations by Müh et
al.\cite{mueh14a}, the results are accurately matching the findings by Jassas et
al.\cite{jass18a}. This agreement is due to the fact that the TrESP approach was
applied in both studies to calculate the excitonic couplings and energies.
Subsequently the absorption spectrum was fitted to determine the site energies.
However, in case of the results by Mascoli et al.\cite{masc20a}, the site
energies differ slightly  from the former results due to the
utilization of the dipole-dipole approximation to determine the excitonic energies
and the corresponding spectrum. As mentioned above, the results based on  DFT
approaches clearly show an overall overestimation of the excitation energy gaps
which is also reflected in a shift in site energies compared to the experimental
outcomes. For this reason, we introduced a common shift for our  site energies 
towards the experimental values and  compared with measurements by Jassas et
al.\cite{jass18a} and Mascoli et al.\cite{masc20a}. The shifted energies are
depicted in the Fig.~S4.

\begin{figure} [tb]
\centering \includegraphics[width=0.75\textwidth]{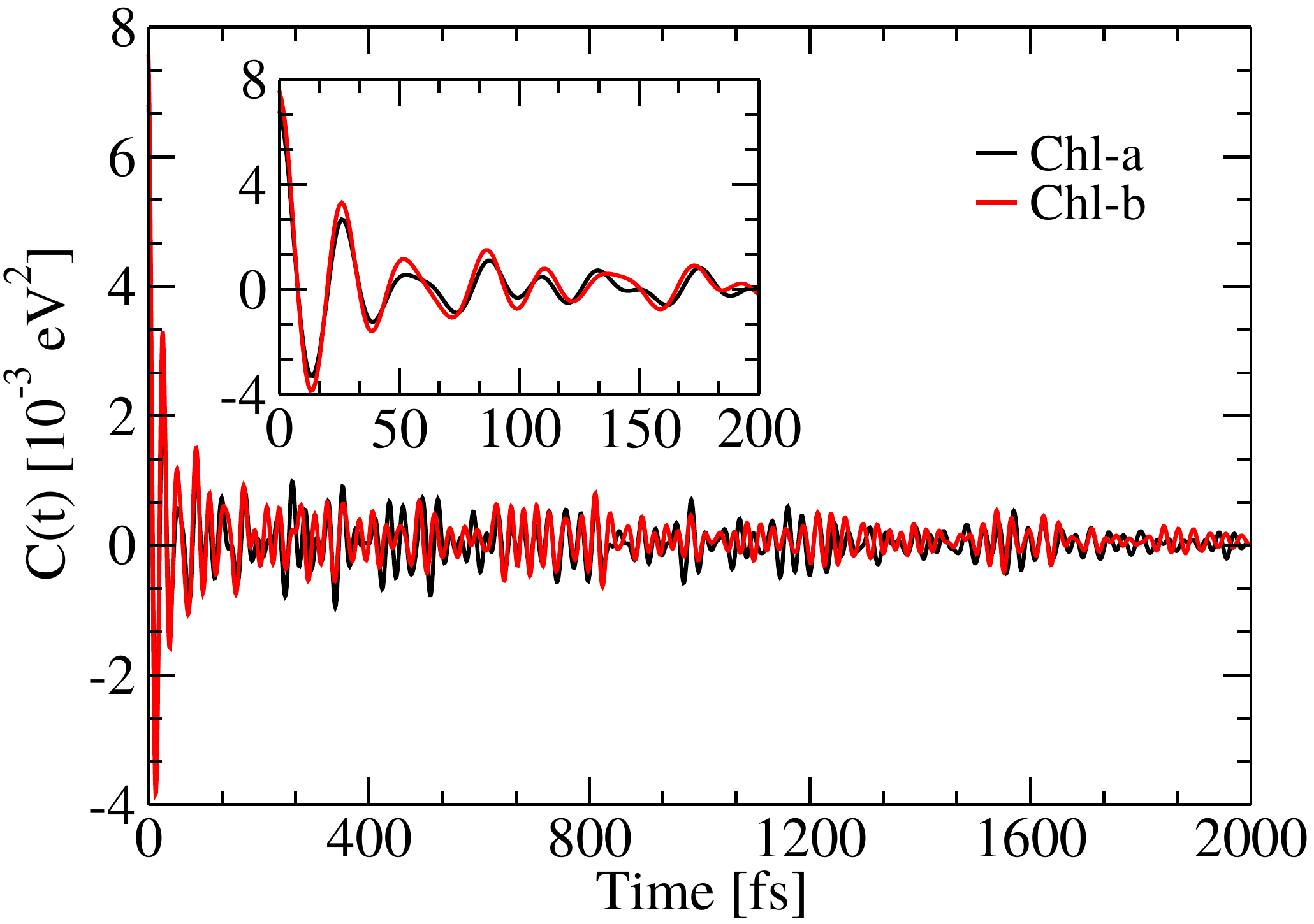}
\caption{\label{fig:auto_CLA_CLB} Average Chl-a and Chl-b autocorrelation function calculated along the two sets of QM/MM MD trajectories and averaged over the pigments of the same time. The inset highlights the first 200~fs.}
\end{figure}

\subsection{Autocorrelation Functions and Spectral Densities} 
\begin{figure} [tb]
\centering \includegraphics[width=0.75\textwidth]{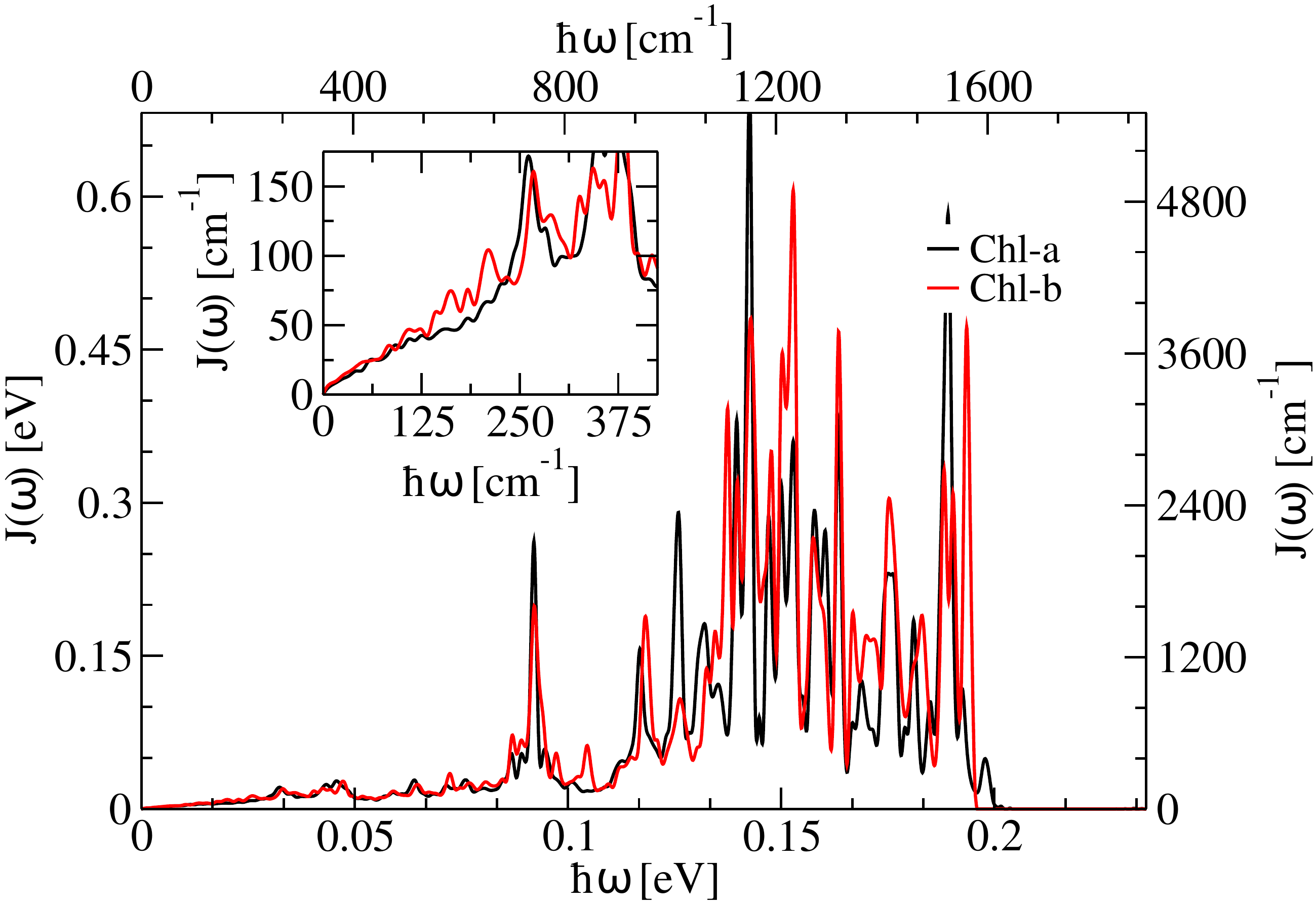}
\caption{\label{fig:spd_CLA_CLB} Spectral densities based on the autocorrelation functions shown in Fig.~\ref{fig:auto_CLA_CLB}.}
\end{figure}

The primary step of calculating spectral densities from site energy trajectories is to compute
the  autocorrelation functions of the site energy fluctuations. Here, we have determined the correlation functions for each Chl-a and Chl-b molecule
separately. The autocorrelation functions averaged over the two 60~ps-long QM/MM MD trajectories and the nine Chl-a and the four Chl-b molecules, respectively,
are depicted in Fig.~\ref{fig:auto_CLA_CLB}. From the figure, it is clear that the Chl-a and Chl-b molecules have very similar correlation functions. This was,
however, to be expected  because of their very similar structures and fluctuations of site energies. The resulting correlation functions show that the shortest  oscillation period is around 22-32 fs which can be attributed to intramolecular collective modes including C=C, C=O and C=N vibrational
stretching in the porphyrin rings. This observation is in line with our previous work on different LH systems based on DFTB-QM/MM MD
simulations \cite{mait20a,mait21a}. The associated spectral densities are shown in Fig.~\ref{fig:spd_CLA_CLB} for the  Chl-a and Chl-b molecules and, as to be
expected, also show very similar lineshapes. The spectral densities of the individual Chl molecules are shown in  Figs.~S5 and S6.

\subsection{Comparison to Experimental Spectral Densities} In experiment,
$\Delta$FLN spectroscopy is used to determine the exciton-phonon and exciton-vibrational coupling of pigment molecules. From the fluorescence
profiles, the corresponding frequencies $\omega$ are extracted together with the respective Huang-Rhys (HR) factors. In a subsequent step, these  HR factors and
the frequencies $\omega$ are used to construct the spectral  densities. Often these spectral densities are given in the form
\begin{equation}
J^{exp} (\omega) = J^{exp}_0 (\omega) + J^{exp}_{vib} (\omega)~,
\label{eq:exp}
\end{equation} 
where $J^{exp}_0 (\omega)$ represents the continuous low frequency  component of the spectral density representing the exciton-phonon coupling, i.e., the coupling to
the protein and solvent environment. This part is estimated as one over-damped Brownian oscillator. The second part, $J^{exp}_{vib} (\omega)$,
describes the high frequency component consisting of more or less separate parts describing intramolecular vibrations of the pigment molecules. Experimentally,
in total 48 vibrational frequencies were extracted for the Chl molecules in the entire PSII system \cite{benn13a} or separately for the LHCII complex \cite{novo04a} which are generally described by under-damped Brownian oscillators. The first part of the spectral density can be modeled by a log-normal expression \cite{kell13a}
\begin{equation}
J^{exp}_0(\omega) = \frac{\hbar \omega S}{\sigma \sqrt{2\pi}} \exp^{-[\ln(\omega / \omega_c)]^2/2\sigma^2} \label{eq:exp-0}
\end{equation}
where the HR factor $S$, the cut-off frequency $\omega_c$ and  the standard deviation $\sigma$ are given in Fig.~3 of Ref.~\citenum{kell13a}. The second
part of the spectral density is specified  by a sum of Lorentzian functions given by
\begin{equation}
J^{exp}_{vib}(\omega) =  \frac{2 \hbar}{\pi} \sum_{k} s_k \omega_k^3 \frac{\gamma_k \omega }{(\omega_k^2 - \omega^2)^2 + \gamma_k^2 \omega^2 }
\label{eq:exp-vib}
\end{equation}
where the HR factors $s_k$ and their corresponding frequencies $\omega_k$ have
been extracted  from  Table~S4 in Ref.~\citenum{masc20a} for the  CP29 complex.
Moreover, the  broadening factor $\gamma_k$ has been chosen to be
$\hbar\gamma_k$ = 7 cm$^{-1}$ (for all $k$) to obtain intensities of the
experimental spectral density peaks similar to those calculated here. This
broadening factor $\gamma_k$ can be chosen freely within certain limits.
Moreover, the HR factors utilized for the CP29 complex are the same as for the
LHCII complex\cite{novo04a} but scaled by a factor in order to reproduce the
fluorescence spectrum \cite{masc20a}.

\begin{figure} [tb]
\centering \includegraphics[width=\textwidth]{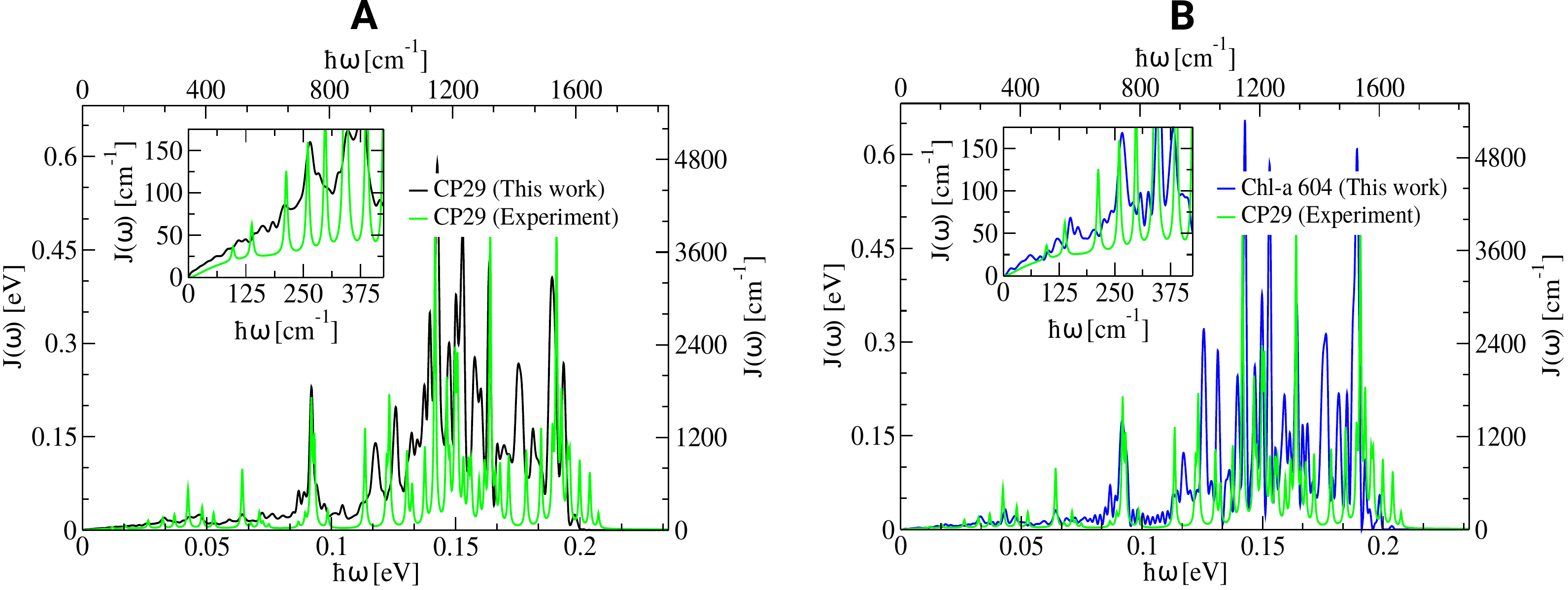}
\caption{\label{fig:spd_exp} A) Average theoretically determined spectral
	density of the CP29 complex in comparison to the experimental spectral densities
	of the CP29 \cite{masc20a}. The inset enhances the low frequency region. B) Same
	as panel A but for the single  pigment Chl-a 604.}
\end{figure} 

The comparison of the theoretically calculated average spectral density of
the CP29 complex with the experimentally determined one are depicted in Fig.~\ref{fig:spd_exp}A. Since the average over all pigments washes out some peaks, here we have also compared the experimental
results with the individual pigment Chl-a 604 (shown in Fig.~\ref{fig:spd_exp}B)
which has the lowest site energy in our calculations. In both plots, the simulated
spectral density for the  CP29 complex shows a remarkable agreement with the
experimental results. The major peaks and their amplitudes are matching well as
shown in Fig.~\ref{fig:spd_exp}. This finding agrees well with our previous
studies for the FMO\cite{mait20a} and the LHCII complexes\cite{mait21a}. In the
low frequency region, however, the amplitude is moderately overestimated. The
reason is not entirely clear but may be due to inaccuracies in the classical
point charges in the QM/MM simulations and the finite length of the correlation
functions since low frequencies correspond to modes with long periods. Further
research in this direction is ongoing.

\subsection{Impact of the Classical Force Fields on Spectral Density in the QM/MM MD Simulation}

\begin{figure} [tb]
\centering \includegraphics[width=0.75\textwidth]{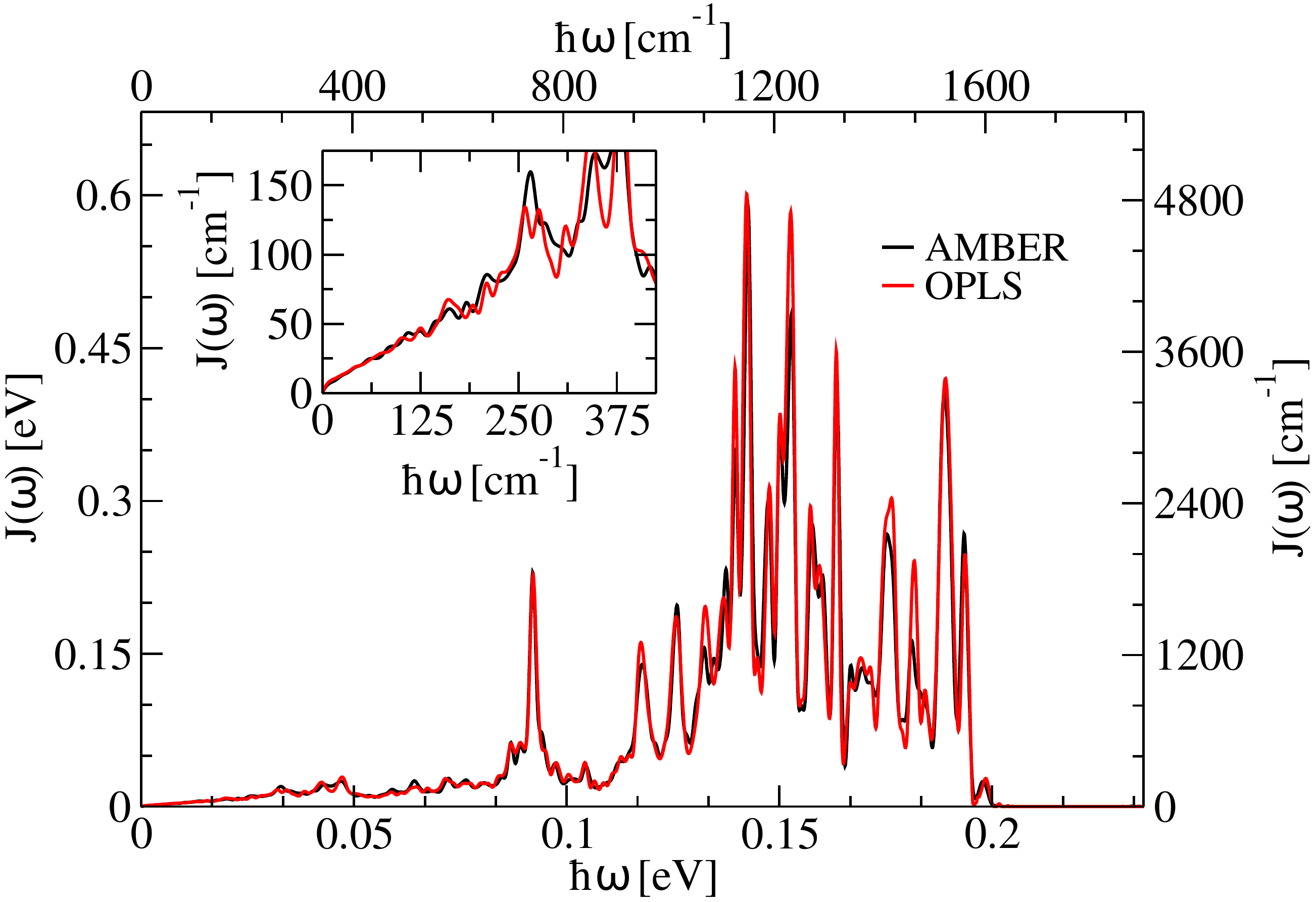}
\caption{\label{fig:spd_amber_opls} Average spectral density of the  CP29 complex based on the  DFTB approach in combination either with the AMBER or the OPLS force field. The inset highlights the low frequency region.}
\end{figure} 
The major peaks in the spectral densities lie in the range from 1030 to 1550
cm$^{-1}$ caused by the fastest oscillations in the  correlation functions.
Here, we would like to point out that in case of spectral densities based on
pure classical force fields, the major peaks were located in the region from
1450 to 1800 cm$^{-1}$ as found in  previous work on bacterial LH
complexes\cite{olbr11b,olbr10a,aght14a,mait20a}. Furthermore, the spectral
densities for Chl molecules based on the semi-empirical PM6 method in a QM/MM MD
dynamics showed a shift in the dominant frequencies by about  100 to 130
cm$^{-1}$ compared to the experimental results\cite{rosn15a}. The inaccurate
positions of the high-amplitude peaks in the spectral densities  arise due to
poor descriptions of the vibrational features using classical force fields or
low-level semi-empirical theories. Using DFTB ground state dynamics together
with the  3OB-f parameter set, however, leads to a proper description of the
most important vibrational frequencies. The choice of the DFTB method rather
than DFT approaches becomes necessary  due to the high numerical cost of the
latter for larger molecules especially when calculating along trajectories. The
low frequency part in the spectral densities is  due to electrostatic
interactions with the environment, i.e., protein as well as  water and ions.
However, different force field sets follow (partially) unlike parametrization
schemes for the  partial charges. Since the partial charges are key ingredients in
electrostatic QM/MM schemes as employed in the present study, we have analyzed
the effect of different force field sets. For the ground state dynamics based
only on classical MD simulations a comparison between  CHARMM and
AMBER-compatible force fields has been performed earlier\cite{chan15a}.
Although, quite some differences between the spectral densities were seen in
that case one has to keep in mind that the parametrization of the pigment
molecules was of large importance. For this reason, in the present case, we have
utilized a setup of the CP29 complex pre-equilibrated using the OPLS force field
\cite{dask18a} and determined the spectral densities in the same way as for the
AMBER force field in a QM/MM fashion. As can be seen in
Fig.~\ref{fig:spd_amber_opls}, the average spectral density based on the OPLS
force field shows an almost identical profile to that one based on the  AMBER
force field in the QM/MM simulations. The positions of the peaks are at the same
for both variants since these  basically rely on the QM part, i.e., the DFTB
approach only which was the same in both variants. More surprisingly, the
contributions in the low frequency part of the spectral density are almost
identical although the environment is described by different force fields, i.e.,
different sets of partial charges. This finding is probably due to the fact that
the partial charges in both both force fields are determined by  fitting of the  electrostatic potentials \cite{jorg88a}. The observed similarity between the QM/MM
simulations in combination with different force fields suggests that both force fields
are equally well suited for the present kind of calculations on LH systems.

\subsection{Comparison of Spectral Density for  other LH Complexes}

\begin{figure} [tb]
\centering \includegraphics[width=0.75\textwidth] {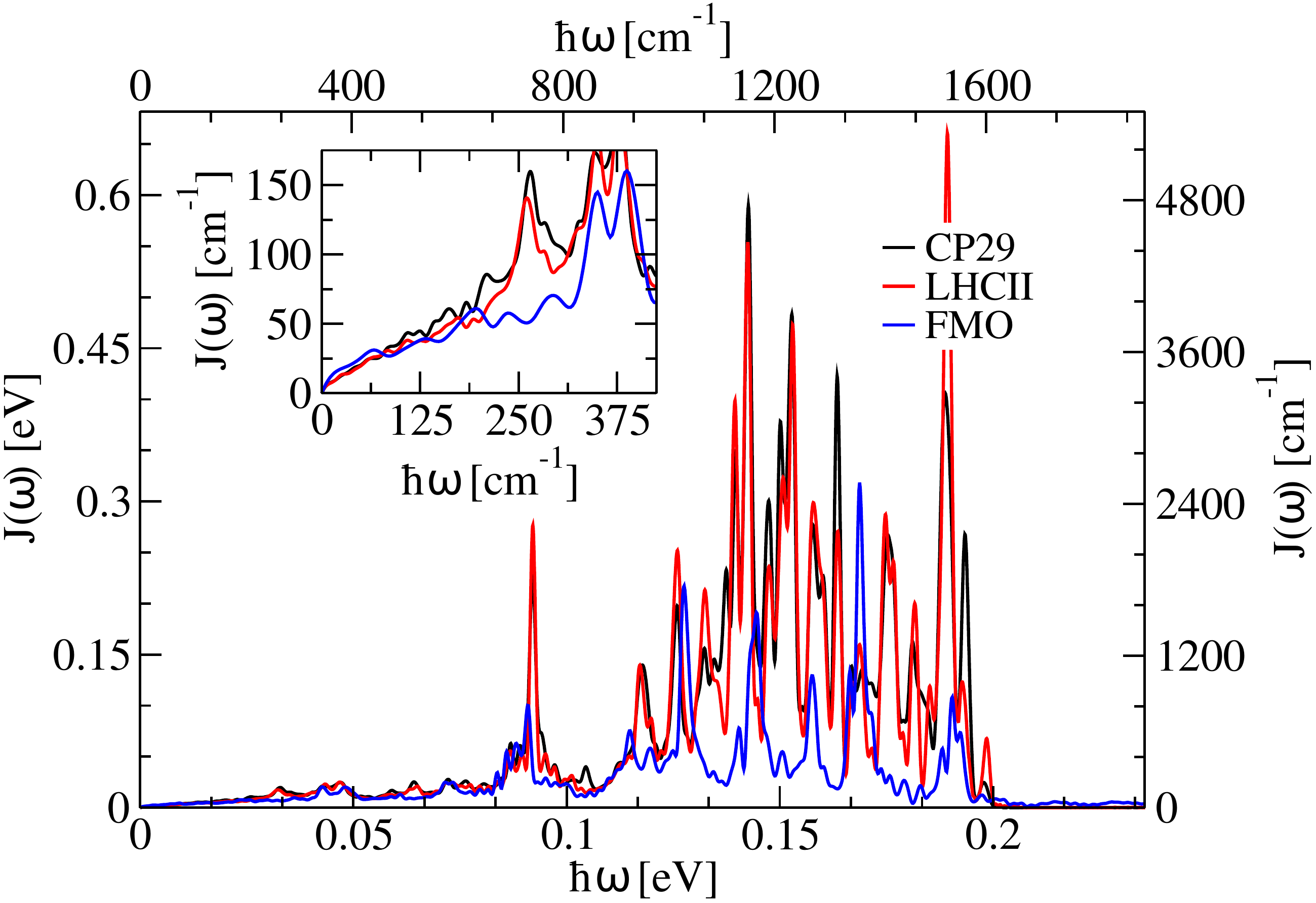}
\caption{\label{fig:spd_fmo_lhcII} Average spectral density of the CP29 complex
	compared  to those from the FMO\cite{mait20a} and the LHCII\cite{mait21a}
	system.}
\end{figure} 
Furthermore, in order to analyze the accuracy and robustness of our method for
LH systems, in Fig.~\ref{fig:spd_fmo_lhcII} we have compared the average
spectral density of the CP29 complex with those of the FMO\cite{mait20a} and
LHCII\cite{mait21a} complexes obtained using the same QM/MM procedure. In case
of the LHCII complex, the Chl-a pool was considered for the whole trimeric
complex\cite{mait21a} whereas for  FMO, all BChl-a pigments have been taken into
account from one monomeric unit of the trimer\cite{mait20a}. As can be seen, the
CP29 and LHCII complexes have almost identical average spectral density
profiles. This is, however, not too surprising since in both  complexes Ch-a and
Chl-b are considered which show almost identical spectral densities as described
earlier. In case of the bacterial FMO complex, however, the amplitudes of the
major peaks are lower compared to those of the plant complexes\cite{mait20a}.
This finding is consistent with the fact that the range of site energy
fluctuations is higher for the Chl molecules in the respective complexes than
those of the BChl pigments in the FMO protein. Likely, the larger values of the
Q$_y$ excitation energies in Chl molecules is one of the fact behind this larger
site energy fluctuations in plant LH complexes, however, further investigation is
required to understand this observation in more detail. Moreover, the number of major peaks
in the FMO complex is less than for the two plant ones which seems to be
surprising since the porphyrin ring of the BChl molecules is more flexible
having one  C=C double bond less than the Chl molecules (see Fig.~S1). This
computational observation is also underlined by experimental measurements  since
in case of the FMO complex 62 vibrational peaks were resolved\cite{raet07a}
whereas for the plant systems 48 peaks were found\cite{novo04a,benn13a}. A
comparison between the  experimental spectral densities of the FMO and CP29
complexes  are shown in the   Fig.~S7. The higher number of vibrational peaks in the
experiments suggests that in our  calculations, some peaks have probably been washed
out during the averaging  procedure over all pigments or that some peaks might have merged
 to form a single peaks with larger width.

\subsection{Excitonic Coupling and Wave Packet Dynamics} 
\begin{figure} [tb]
\centering \includegraphics[width=0.70\textwidth]{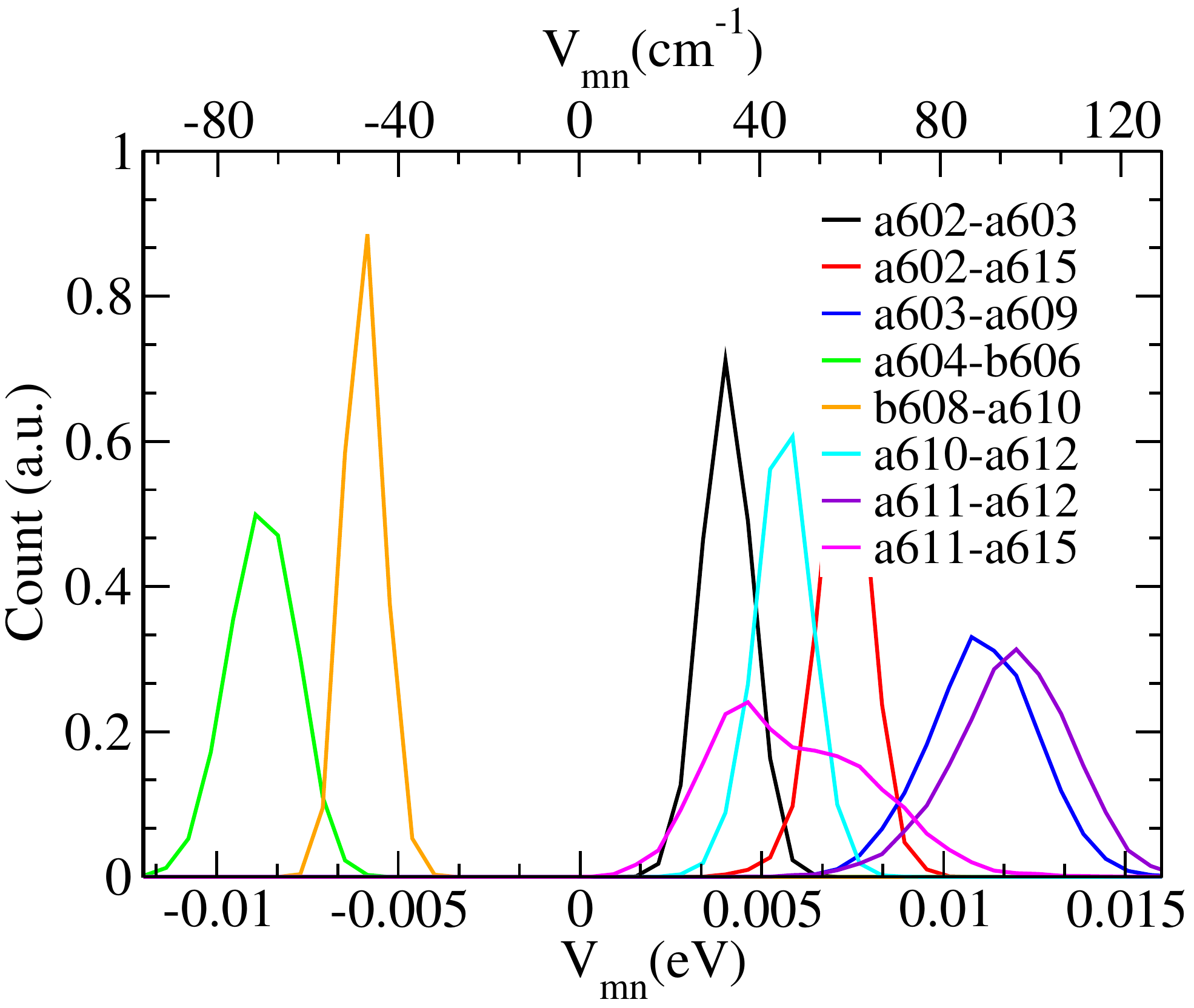}
\caption{\label{fig:TrESP_Coupling} Distribution of excitonic coupling values for the pigment pairs whose average absolute values are above 30 cm$^{-1}$.}
\end{figure}

DFTB-QM/MM MD 
simulations are still computationally expensive when one wants to treat  several
pigments at a time. Therefore,  we have constructed the time-averaged
Hamiltonian based on the coupling values calculated from the 200~ns classical MD
simulation and the site energies based on the 1~ns QM/MM MD trajectory (see
Table.~\ref{table:hamil}). As described in the methods section, the excitonic 
couplings between all Chl pigments of the CP29 complex were calculated using the
TrESP approach. A total of 10,000 frames from the 200~ns long classical MD
trajectory were employed to calculate the coupling values. The distribution of
larger coupling values are represented in Fig.~\ref{fig:TrESP_Coupling}. Most of
the larger coupling values originate from the Chl-a pairs as a consequence of
their spatial proximity. In case of the Chl-b chromophores, pigments 606 and 608
are participating in stronger couplings with Chl-b 606 shows the highest
coupling value to neighboring pigments. Once the excitonic couplings have been
determined, the system Hamiltonian can be established based on
Eq.~\ref{eq:Hamil}. This time-independent Hamiltonian together with the spectral
densities obtained in the present study can serve as a starting point for future
calculations  using  density matrix approaches.

\begin{table}[H]
\centering \caption{Time-averaged system Hamiltonian of the CP29 complex based
	on the exciton coupling values along the 200~ns classical MD trajectory and the
	site energies from the  1~ns QM/MM MD simulation. The site energies 
	and   couplings (in cm$^{-1})$ with absolute values  above 30 cm$^{-1}$ are shown in bold.}
\scalebox{0.75}{
\begin{tabular}{|c|c|c|c|c|c|c|c|c|c|c|c|c|c|}
\hline
& \textbf{a602} & \textbf{a603} & \textbf{a604} & \textbf{b606} & \textbf{b607} & \textbf{b608} & \textbf{a609} & \textbf{a610} & \textbf{a611} & \textbf{a612} & \textbf{a613} & \textbf{b614} & \textbf{a615} \\ 
\hline
\textbf{a602}  &  \textbf{16594}  &  \textbf{30.08}  &  4.23  & -3.33  &  -1.68  &  3.85  &  -23.65 &  -5.75  & -1.26  &  7.32  & -1.98  & -0.58  &  \textbf{56.03} \\
\textbf{a603}  &  \textbf{30.08}  &  \textbf{16617}  & -1.06  &  2.97  & -9.64   & -2.96  &  \textbf{86.80}  &  6.50   & -0.67  & -2.06  &  0.66  &  2.92  & -3.65  \\
\textbf{a604}  &  4.23   & -1.06   &  \textbf{16408} & \textbf{-72.97} & -2.37   &  2.05  & -0.67   & -1.58   & -1.97  &  0.77  &  0.52  &  1.95  & -2.18  \\
\textbf{b606}  & -3.33   &  2.97   & \textbf{-72.97} &  \textbf{16796} &  1.73   & -1.89  & -5.14   &  1.18   &  1.50  & -1.11  & -0.73  & -0.82  &  1.33  \\
\textbf{b607}  & -1.68   & -9.64   & -2.37  &  1.73  &  \textbf{16997}  & -2.12  &  8.71   & -1.87   &  1.08  & -0.13  &  1.26  & -1.71  &  0.23  \\
\textbf{b608}  &  3.85   & -2.96   &  2.05  & -1.89  & -2.12   &  \textbf{17442} & -22.17  & \textbf{-50.39}  & -3.48  &  1.51  &  1.42  &  0.62  & -3.00  \\
\textbf{a609}  & -23.65  &  \textbf{86.80}  & -0.67  & -5.14  &  8.71   & -22.17 &  \textbf{16612}  & -0.94   &  3.13  & -0.26  & -2.15  & -1.28  &  4.62  \\
\textbf{a610}  & -5.75   &  6.50   & -1.58  &  1.18  & -1.87   & \textbf{-50.39} & -0.94   &  \textbf{16360}  & -27.43 &  \textbf{42.57 }&  5.23  &  0.12  & -2.85  \\
\textbf{a611}  & -1.26   & -0.67   & -1.97  &  1.50  &  1.08   & -3.48  &  3.13   & -27.43  &  \textbf{16462} &  \textbf{93.09} & -4.11  &  1.18  &  \textbf{44.99}  \\
\textbf{a612}  &  7.32   & -2.06   & 0.77   & -1.11  & -0.13   &  1.51  & -0.26   &  \textbf{42.57}  &  \textbf{93.09} &  \textbf{16617} & -6.078 & -1.50  & -3.82   \\
\textbf{a613}  & -1.98   &  0.66   & 0.52   & -0.73  &  1.26   &  1.42  & -2.15   &  5.23   & -4.11  &  -6.08 &  \textbf{16409} &  7.38  &  6.48   \\
\textbf{b614}  & -0.58   &  2.92   & 1.95   & -0.82  & -1.71   &  0.62  & -1.28   &  0.12   &  1.18  &  -1.50 &  7.38  &  \textbf{17030} & 10.65   \\
\textbf{a615}  &  \textbf{56.03}  & -3.65   & -2.18  & 1.33   &  0.23   & -3.00  & 4.62    & -2.85   & \textbf{44.99}  & -3.82  & 6.48   & 10.65  & \textbf{16537}   \\
\hline
\end{tabular}}
\label{table:hamil} \\
\end{table}

As accurate density matrix calculations remain numerically  expensive for spectral
densities as the ones derived here,  we have performed ensemble-averaged
wave-packet dynamics within the Ehrenfest approach (without back reaction of the
thermal bath). For certain parameter regimes, the Ehrenfest wave packet approach has been
shown to yield the same results as accurate density matrix calculations properly
representing the dephasing  but not the relaxation \cite{aght12a}. In the wave packet-based scheme, the time-dependent
Schrödinger equation needs to be solved  for the time-dependent system
Hamiltonian
\begin{equation}
i\hbar \frac{\partial| \Psi_S(t) \rangle}{\partial t} = H_S(t) |\Psi_S(t)\rangle \label{eq:SE}
\end{equation}
where $|\Psi_S(t)\rangle$ denotes an excitonic state in the single-exciton manifold. This state can be expanded in terms of time-independent states $\ket{\alpha}$
\begin{equation}
 |\Psi_S(t)\rangle = \sum_{\alpha}c_{\alpha}(t)\ket{\alpha}
\end{equation}
with time-dependent coefficients  $c_{\alpha}(t)$. Moreover, the excitonic states $\ket{\alpha}$ can be written in terms of the site-local states  $\ket{m}$ as
\begin{equation}
\ket{\alpha} = \sum_{m} c_{m}^{\alpha} \ket{m}~.
\label{eq:basis}
\end{equation}
Combining these equations, the probability density of finding an exciton on an individual pigment site $m$ is given by
\begin{equation}
P_m(t) =  |\langle m |\Psi_S(t) \rangle|^2 = |\sum_{\alpha} c_m^\alpha~.
c_{\alpha}(t)|^2 \label{eq:prob}
\end{equation}
The dynamics of the probability density $P_m(t)$ shows how an exciton can propagate from pigment to pigment and spread at the same time.  To obtain a meaningful exciton transfer
dynamics, the solution of the time-dependent Schrödinger equation needs to be repeated many times form different starting point along the trajectory to
obtain ensemble-averaged results.

\begin{figure} [tb]
\centering \includegraphics[width=\textwidth]{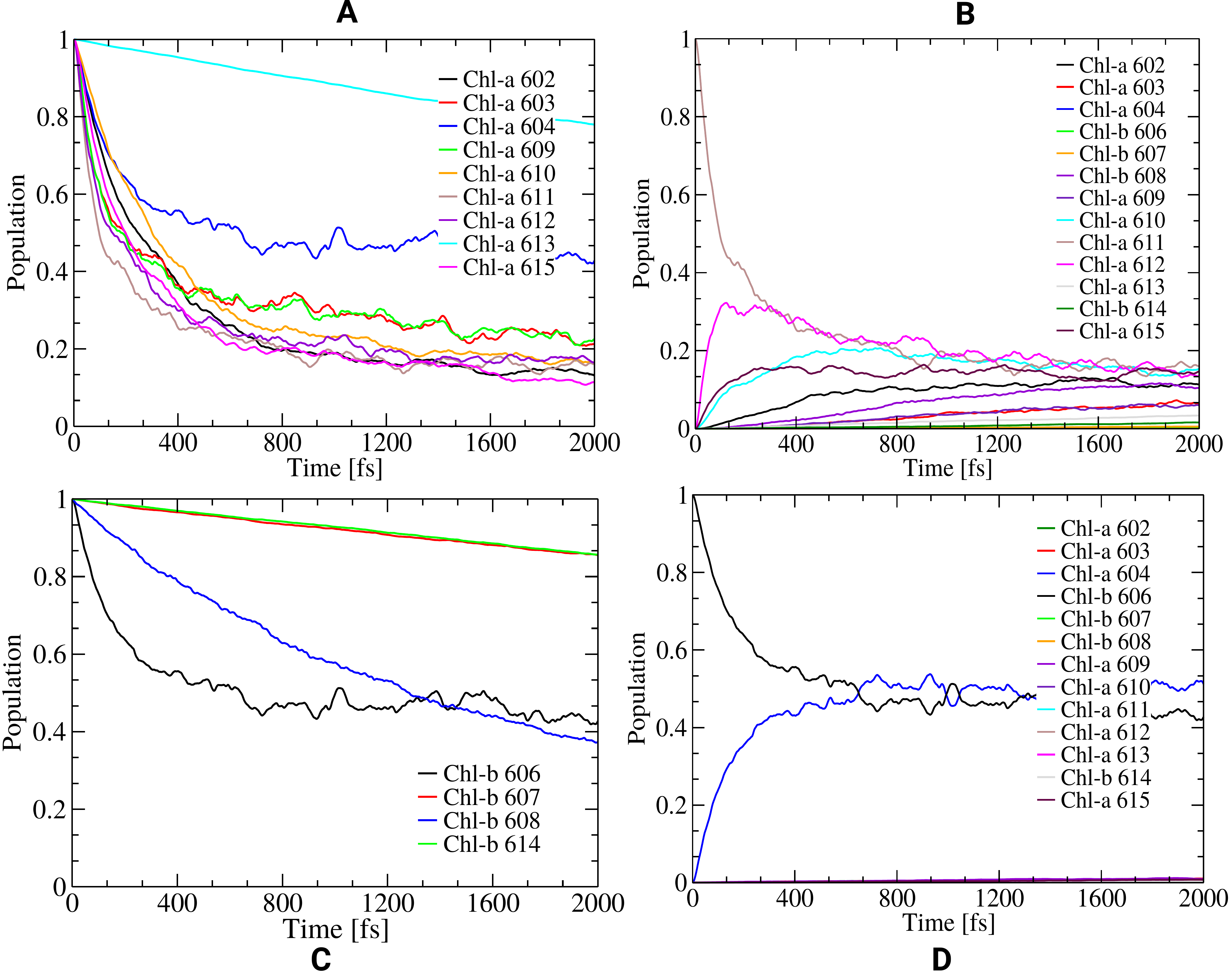}
\caption{\label{fig:pop} A) Decay of the exciton population from different initially excited Chl-a pigments. B) Population transfer from an initially excited Chl-a 611 pigment to the other pigment molecules inside the  CP29 complex. C) Same as panel A but for the Chl-b pigments. D) Same as panel B but for the initially excited Chl-b 606 pigment.}
\end{figure} 
In case of the CP29 complex, we have taken the  average coupling values
calculated from the 200~ns classical MD simulation together with the site
energies obtained from the first set of the 60~ps QM/MM MD trajectories to
propagate the wave packet dynamics. This leads to a realistic time-dependent
Hamiltonian for the 60~ps trajectory since the coupling fluctuations  hardly
impact the exciton propagation as shown in earlier studies
\cite{aght17a,mait20a}. Furthermore, for the averaging procedure we have assumed
that the temporal site energy fluctuations are basically uncorrelated after
500~fs (see also Fig.~\ref{eq:acf}). Thus, every 500~fs a new starting point in
the trajectory is used for the averaging procedure using a sliding window
technique. In our calculations, we initially excite a single pigment and then
monitor the propagation of the exciton wave function towards the other pigments
inside the CP29 complex. Each of the 13 Chl pigments was excited individually
and the exciton distributions were captured for each case. To this end,
Fig.~\ref{fig:pop}A shows the decay of an excitation from individually excited
Chl-a pigments. The pigment  Chl-a 611 shows the fastest decay since it has the
largest excitonic coupling values to its neighbors  whereas for  Chl-a 613 the
decay is very slow because of its weak interpigment couplings. In
Fig.~\ref{fig:pop}B we show the case of Chl-a 611 separately including the
populations of the other chlorophylls to which the excitation energy is
transferred. The population leaves pigment Chl-a 611 exponentially fast and
moves to neighboring pigments within a few 100 fs. Since Chl-a 612 is strongly
coupled to Chl-a 611, its population gain is very fast within the first 400~fs
before the excitation energy is transferred further. At this point one has to
keep in mind that due to the implicit high-temperature approximation in the
Ehrenfest approach, the scheme will not lead to a proper thermodynamic
equilibrium state but rather to an equal population of all sites. Moreover, it
should be mentioned that the population curves become smoother within increasing
the number of samples in the averaging procedure which is limited here due to
the finite length of the QM/MM trajectory.

Similar to the  Chl-a pigments, the transfer dynamics for the Chl-b molecules is
shown in Figs.~\ref{fig:pop}C and D. Since the pigments  Chl-b 607 and 614 have
only  weak coupling values to the other chlorophylls, the exciton transfer is
expected to be slow from these sites. As one can see in Fig.~\ref{fig:pop}C,
more than 80\% of the population is still present on these pigments if they were
initially excited. The coupling values of  pigment Chl-b 608 are slightly higher
which is also visible in  the respective exciton transfer time. However, Chl-b
606 is strongly coupled to its neighboring pigment Chl-a 604 and within about
500~fs, almost the full population is equally shared between these pigments.
Moreover, the dynamics continues for a longer time more than 1.5~ps where the
population gain of the other pigments is close to zero. 

\section{Conclusions} The molecular details and the mechanism of the pH-induced
photoprotective mechanism in higher plants has raised significant interest in
recent years\cite{ruba07a,ruba12a,chme16a,dask19b}. It has been proposed  that
under excess sunlight an enhanced pH gradient is established across the
thylakoid membrane  embedding the PSII complex which induces conformational
changes in the associated proteins in order to release the extra energy as heat.
In this process, the major antenna complex LHCII has been identified as one of
the key players. Furthermore, recent studies show that the minor antenna complex
CP29 which acts as a bridge between  LHCII and the reaction center is also
involved in this quenching mechanism\cite{nico19a,guar20a,tian19a}. In order to
study the energy transfer dynamics involved in this system, the bath-induced
spectral density is one of the key ingredients in tight-binding descriptions of
the underlying processes.

Earlier theoretical approaches for calculating spectral densities were based on
classical MD simulations for the ground state dynamics followed by excited state
calculations. In these calculations, the  high frequency regions of these
spectral densities were rather inaccurately described since standard classical force
fields cannot properly describe the intramolecular vibrational dynamics of the
respective pigments. To avoid this issue, a QM/MM MD ground state dynamics has
to be performed. Using DFT approaches with a reasonable functional and basis set
is numerically quite expensive. To this end, we performed the ground state QM/MM
MD dynamics followed by  excited state calculations within the numerically
efficient DFTB formalism. This procedure  yielded spectral densities with a
remarkable accuracy compared to experimental findings.  In this study, we have
extended this multiscale approach to the CP29 minor antenna complex containing
Chl-a and Chl-b molecules as primary pigments.

The trend of the site energy ladder is in good agreement with other computed and
measured results while the  values of the site energies are
overestimated due to the well-known issues of DFT approaches with overestimating
excitation energy gaps \cite{mait20a, mait21a}. For this reason, we introduced a
common shift for the present site energies towards the experimental energies to match the average experimental
results. Moreover, as expected, the average site energies based on  TD-LC-DFTB
calculations along  DFTB-QM/MM MD trajectories  for the Chl-b molecules in the
CP29 complex are slightly blue shifted compared to those of the Chl-a
chromophores. However, in the energy gap autocorrelation functions and spectral
densities, such a difference is not visible since both Chl molecules produce
very similar correlation functions and spectral densities. Moreover, the
accuracy of the present method can also be judged by comparing with  spectral
densities obtained from experiments. Most of the major peaks are found at the
same frequencies and the intensities show a very good agreement with 
experimental results. In the present findings the lower
frequency peaks are, however, moderately overestimated which needs to be further
analyzed.

Moreover, in order to check the robustness and reliability of the present
scheme, two different variants of classical force fields, i.e., AMBER and OPLS,
were employed together  with the DFTB approach in a QM/MM fashion. The
description of the environment with the two different force fields resulted in
almost indistinguishable spectral densities which supports the fact that either
of the two force fields might be used  when  modeling  LH complexes within a
QM/MM framework. Moreover, based on our previous study on the FMO complex of
green sulfur bacteria, the range of the site energy fluctuations of BChl
pigments is noticeable smaller than that of the Chl molecule calculated in the
present study  and in a previous one for the LHCII complex \cite{mait21a}. This fact  is also
reflected in the amplitudes of the spectral density peaks which are lower for
the FMO complex when compare to those of the two plant systems. In experimental
measurements, the spectral density of the FMO complex shows more peaks since it
contains one C=C double bond less in the  Mg-porphyrin rings compared to the
bacterial systems. However, in our theoretical calculations, we found  a smaller
number of peaks since some of them seem to be merged to form wider peaks.

In addition to determining site energies and spectral densities, we have also
calculated the  excitonic couplings based on the TrESP approach and constructed
a time-averaged system Hamiltonian. Since the  QM/MM MD simulation numerically
is still expensive when talking several pigments into account simultaneously,
the couplings were determined from a 200~ns classical MD trajectory and combined
with the  site energies from the 1~ns QM/MM MD trajectory. This mixed approach
is reasonable as the coupling fluctuations are less significant during the
energy transfer dynamics \cite{aght17a, mait20a}. A time-averaged version of the
Hamiltonian together with the obtained spectral densities can possibly be
employed in  future density matrix-based studies. In the present study, however,
the time-averaged coupling values together with the time-dependent site energies
from the first set of 60 ps QM/MM MD trajectories have been used in an
ensemble-averaged  wave packet scheme. Using such an approach, one can analyze 
how  excitons are transferred between the individual pigments and how they
spread within the system.

Recently, we have established a  multiscale protocol with the numerically
efficient DFTB method which can accurately describe   spectral densities which
are key quantities for  modeling of photochemical
processes\cite{mait20a,mait21a}.  In the present study, the approach has been
tested once more for an important plant LH system and it has been shown that the
scheme produces reproducible, robust and reliable results based on different
simulations and various LH complexes. To this end, the  multiscale protocol was
utilized for  the CP29 complex which  recently has gained quite some interest
because of its active participation in the non-photochemical
quenching\cite{son19a,masc20a}. The outcome of the present calculations are the
site energy, couplings and most importantly the spectral densities. Moreover, we
have also presented a realistic time-averaged Hamiltonian and obtained results
for the exciton dynamics based on a  time-dependent Hamiltonian. In conclusion,
the present study has shed light on the electronic properties of the pigment
molecules in the CP29 complex as well as on the exciton-phonon and
exciton-vibrational couplings within this pigment-protein aggregate. This
molecular-level insight can be used in future investigations of this
biologically relevant complex or of the whole PSII machinery. Simulations of
entire bacterial photosynthetic organelles became possible recently
\cite{sing19a} while similar simulations of larger aggregates of plant
pigment-protein complexes are yet to come.

\section*{Acknowledgments} The authors acknowledge support by the DFG through  grant KL-1299/18-1 as well as through the Research Training Group 2247
``Quantum Mechanical Materials Modelling''.

\section*{Data Availability}
The data that support the findings of this study are available from the corresponding author upon reasonable request.

%\bibliography{ukleine}

\begin{mcitethebibliography}{91}
\providecommand*\natexlab[1]{#1}
\providecommand*\mciteSetBstSublistMode[1]{}
\providecommand*\mciteSetBstMaxWidthForm[2]{}
\providecommand*\mciteBstWouldAddEndPuncttrue
  {\def\EndOfBibitem{\unskip.}}
\providecommand*\mciteBstWouldAddEndPunctfalse
  {\let\EndOfBibitem\relax}
\providecommand*\mciteSetBstMidEndSepPunct[3]{}
\providecommand*\mciteSetBstSublistLabelBeginEnd[3]{}
\providecommand*\EndOfBibitem{}
\mciteSetBstSublistMode{f}
\mciteSetBstMaxWidthForm{subitem}{(\alph{mcitesubitemcount})}
\mciteSetBstSublistLabelBeginEnd
  {\mcitemaxwidthsubitemform\space}
  {\relax}
  {\relax}

\bibitem[Blankenship(2014)]{blan14a}
Blankenship,~R.~E. \emph{Molecular Mechanisms of Photosynthesis}, 2nd ed.;
  Wiley, 2014\relax
\mciteBstWouldAddEndPuncttrue
\mciteSetBstMidEndSepPunct{\mcitedefaultmidpunct}
{\mcitedefaultendpunct}{\mcitedefaultseppunct}\relax
\EndOfBibitem
\bibitem[{E}ngel \latin{et~al.}(2007){E}ngel, {C}alhoun, {R}ead, {A}hn,
  {M}ancal, {C}heng, {B}lankenship, and {F}leming]{enge07a}
{E}ngel,~G.~S.; {C}alhoun,~T.~R.; {R}ead,~E.~L.; {A}hn,~T.~K.; {M}ancal,~T.;
  {C}heng,~Y.~C.; {B}lankenship,~R.~E.; {F}leming,~G.~R. {E}vidence for
  {W}avelike {E}nergy {T}ransfer {T}hrough {Q}uantum {C}oherence in
  {P}hotosynthetic {S}ystems. \emph{{N}ature} \textbf{2007}, \emph{446},
  782--786\relax
\mciteBstWouldAddEndPuncttrue
\mciteSetBstMidEndSepPunct{\mcitedefaultmidpunct}
{\mcitedefaultendpunct}{\mcitedefaultseppunct}\relax
\EndOfBibitem
\bibitem[{C}ollini and {S}choles(2009){C}ollini, and {S}choles]{coll09a}
{C}ollini,~E.; {S}choles,~G.~D. {C}oherent {I}ntrachain {E}nergy {M}igration in
  a {C}onjugated {P}olymer at {R}oom {T}emperature. \emph{{S}cience}
  \textbf{2009}, \emph{323}, 369--373\relax
\mciteBstWouldAddEndPuncttrue
\mciteSetBstMidEndSepPunct{\mcitedefaultmidpunct}
{\mcitedefaultendpunct}{\mcitedefaultseppunct}\relax
\EndOfBibitem
\bibitem[{C}ollini \latin{et~al.}(2010){C}ollini, {W}ong, {W}ilk, {C}urmi,
  {B}rumer, and {S}choles]{coll10a}
{C}ollini,~E.; {W}ong,~C.~Y.; {W}ilk,~K.~E.; {C}urmi,~P. M.~G.; {B}rumer,~P.;
  {S}choles,~G.~D. {C}oherently {W}ired {L}ight-harvesting in {P}hotosynthetic
  {M}arine {A}lgae at {A}mbient {T}emperature. \emph{{N}ature} \textbf{2010},
  \emph{463}, 644--647\relax
\mciteBstWouldAddEndPuncttrue
\mciteSetBstMidEndSepPunct{\mcitedefaultmidpunct}
{\mcitedefaultendpunct}{\mcitedefaultseppunct}\relax
\EndOfBibitem
\bibitem[{P}anitchayangkoon \latin{et~al.}(2010){P}anitchayangkoon, {H}ayes,
  {F}ransted, {C}aram, {H}arel, {W}en, {B}lankenship, and {E}ngel]{pani10a}
{P}anitchayangkoon,~G.; {H}ayes,~D.; {F}ransted,~K.~A.; {C}aram,~J.~R.;
  {H}arel,~E.; {W}en,~J.; {B}lankenship,~R.~E.; {E}ngel,~G.~S. {L}ong-{L}ived
  {Q}uantum {C}oherence in {P}hotosynthetic {C}omplexes at {P}hysiological
  {T}emperature. \emph{{P}roc. {N}atl. {A}cad. {S}ci. {USA}} \textbf{2010},
  \emph{107}, 12766--12770\relax
\mciteBstWouldAddEndPuncttrue
\mciteSetBstMidEndSepPunct{\mcitedefaultmidpunct}
{\mcitedefaultendpunct}{\mcitedefaultseppunct}\relax
\EndOfBibitem
\bibitem[Cao \latin{et~al.}(2020)Cao, Cogdell, Coker, Duan, Hauer,
  Kleinekath{\"{o}}fer, Jansen, Man{\v{c}}al, Miller, Ogilvie, Prokhorenko,
  Renger, Tan, Tempelaar, Thorwart, Thyrhaug, Westenhoff, and
  Zigmantas]{cao20a}
Cao,~J. \latin{et~al.}  Quantum Biology Revisited. \emph{Sci. Adv.}
  \textbf{2020}, \emph{6}, eaaz4888\relax
\mciteBstWouldAddEndPuncttrue
\mciteSetBstMidEndSepPunct{\mcitedefaultmidpunct}
{\mcitedefaultendpunct}{\mcitedefaultseppunct}\relax
\EndOfBibitem
\bibitem[Duan \latin{et~al.}(2017)Duan, Prokhorenko, Cogdell, Ashraf, Stevens,
  Thorwart, and Miller]{duan17b}
Duan,~H.-G.; Prokhorenko,~V.~I.; Cogdell,~R.~J.; Ashraf,~K.; Stevens,~A.~L.;
  Thorwart,~M.; Miller,~R. J.~D. Nature Does Not Rely on Long-Lived Electronic
  Quantum Coherence for Photosynthetic Energy Transfer. \emph{Proc Natl Acad
  Sci USA} \textbf{2017}, \emph{114}, 8493\relax
\mciteBstWouldAddEndPuncttrue
\mciteSetBstMidEndSepPunct{\mcitedefaultmidpunct}
{\mcitedefaultendpunct}{\mcitedefaultseppunct}\relax
\EndOfBibitem
\bibitem[Thyrhaug \latin{et~al.}(2018)Thyrhaug, Tempelaar, Alcocer,
  {{\v{Z}}}{\'{i}}dek, B{\'{i}}na, Knoester, Jansen, and Zigmantas]{thyr18a}
Thyrhaug,~E.; Tempelaar,~R.; Alcocer,~M. J.~P.; {{\v{Z}}}{\'{i}}dek,~K.;
  B{\'{i}}na,~D.; Knoester,~J.; Jansen,~T. L.~C.; Zigmantas,~D. Identification
  And Characterization Of Diverse Coherences In The Fenna-Matthews-Olson
  Complex. \emph{Nat. Chem.} \textbf{2018}, \emph{10}, 780--786\relax
\mciteBstWouldAddEndPuncttrue
\mciteSetBstMidEndSepPunct{\mcitedefaultmidpunct}
{\mcitedefaultendpunct}{\mcitedefaultseppunct}\relax
\EndOfBibitem
\bibitem[Ruban \latin{et~al.}(2007)Ruban, Berera, Ilioaia, Van~Stokkum, Kennis,
  Pascal, Van~Amerongen, Robert, Horton, and Van~Grondelle]{ruba07a}
Ruban,~A.~V.; Berera,~R.; Ilioaia,~C.; Van~Stokkum,~I.~H.; Kennis,~J.~T.;
  Pascal,~A.~A.; Van~Amerongen,~H.; Robert,~B.; Horton,~P.; Van~Grondelle,~R.
  Identification of a Mechanism of Photoprotective Energy Dissipation in Higher
  Plants. \emph{Nature} \textbf{2007}, \emph{450}, 575\relax
\mciteBstWouldAddEndPuncttrue
\mciteSetBstMidEndSepPunct{\mcitedefaultmidpunct}
{\mcitedefaultendpunct}{\mcitedefaultseppunct}\relax
\EndOfBibitem
\bibitem[Ruban \latin{et~al.}(2012)Ruban, Johnson, and Duffy]{ruba12a}
Ruban,~A.~V.; Johnson,~M.~P.; Duffy,~C. D.~P. The Photoprotective Molecular
  Switch in the Photosystem {II} Antenna. \emph{Biochim. Biophys. Acta (BBA) -
  Bioenergetics} \textbf{2012}, \emph{1817}, 167--181\relax
\mciteBstWouldAddEndPuncttrue
\mciteSetBstMidEndSepPunct{\mcitedefaultmidpunct}
{\mcitedefaultendpunct}{\mcitedefaultseppunct}\relax
\EndOfBibitem
\bibitem[Chmeliov \latin{et~al.}(2016)Chmeliov, Gelzinis, Songaila, Augulis,
  Duffy, Ruban, and Valkunas]{chme16a}
Chmeliov,~J.; Gelzinis,~A.; Songaila,~E.; Augulis,~R.; Duffy,~C. D.~P.;
  Ruban,~A.~V.; Valkunas,~L. The Nature of Self-Regulation in Photosynthetic
  Light-Harvesting Antenna. \emph{Nat. Plants} \textbf{2016}, \emph{2},
  16045\relax
\mciteBstWouldAddEndPuncttrue
\mciteSetBstMidEndSepPunct{\mcitedefaultmidpunct}
{\mcitedefaultendpunct}{\mcitedefaultseppunct}\relax
\EndOfBibitem
\bibitem[Tian \latin{et~al.}(2019)Tian, Nawrocki, Liu, Polukhina, Van~Stokkum,
  and Croce]{tian19a}
Tian,~L.; Nawrocki,~W.~J.; Liu,~X.; Polukhina,~I.; Van~Stokkum,~I.~H.;
  Croce,~R. Ph Dependence, Kinetics and Light-Harvesting Regulation of
  Nonphotochemical Quenching in Chlamydomonas. \emph{Proc. Nat. Acad. Sci.}
  \textbf{2019}, \emph{116}, 8320--8325\relax
\mciteBstWouldAddEndPuncttrue
\mciteSetBstMidEndSepPunct{\mcitedefaultmidpunct}
{\mcitedefaultendpunct}{\mcitedefaultseppunct}\relax
\EndOfBibitem
\bibitem[Nicol \latin{et~al.}(2019)Nicol, Nawrocki, and Croce]{nico19a}
Nicol,~L.; Nawrocki,~W.~J.; Croce,~R. Disentangling the Sites of
  Non-Photochemical Quenching in Vascular Plants. \emph{Nat. Plants}
  \textbf{2019}, \emph{5}, 1177--1183\relax
\mciteBstWouldAddEndPuncttrue
\mciteSetBstMidEndSepPunct{\mcitedefaultmidpunct}
{\mcitedefaultendpunct}{\mcitedefaultseppunct}\relax
\EndOfBibitem
\bibitem[Buck \latin{et~al.}(2019)Buck, Sherman, B{\'a}rtulos, Serif, Halder,
  Henkel, Falciatore, Lavaud, Gorbunov, Kroth, \latin{et~al.} others]{buck19a}
Buck,~J.~M.; Sherman,~J.; B{\'a}rtulos,~C.~R.; Serif,~M.; Halder,~M.;
  Henkel,~J.; Falciatore,~A.; Lavaud,~J.; Gorbunov,~M.~Y.; Kroth,~P.~G.,
  \latin{et~al.}  Lhcx Proteins Provide Photoprotection via Thermal Dissipation
  of Absorbed Light in the Diatom Phaeodactylum Tricornutum. \emph{Nat. Comm.}
  \textbf{2019}, \emph{10}, 1--12\relax
\mciteBstWouldAddEndPuncttrue
\mciteSetBstMidEndSepPunct{\mcitedefaultmidpunct}
{\mcitedefaultendpunct}{\mcitedefaultseppunct}\relax
\EndOfBibitem
\bibitem[de~la Cruz~Valbuena \latin{et~al.}(2019)de~la Cruz~Valbuena,
  VA~Camargo, Borrego-Varillas, Perozeni, D’Andrea, Ballottari, and
  Cerullo]{dela19a}
de~la Cruz~Valbuena,~G.; VA~Camargo,~F.; Borrego-Varillas,~R.; Perozeni,~F.;
  D’Andrea,~C.; Ballottari,~M.; Cerullo,~G. Molecular Mechanisms of
  Nonphotochemical Quenching in the {LHCSR3} Protein of {\emph Chlamydomonas
  reinhardtii}. \emph{J. Phys. Chem. Lett.} \textbf{2019}, \emph{10},
  2500--2505\relax
\mciteBstWouldAddEndPuncttrue
\mciteSetBstMidEndSepPunct{\mcitedefaultmidpunct}
{\mcitedefaultendpunct}{\mcitedefaultseppunct}\relax
\EndOfBibitem
\bibitem[Li \latin{et~al.}(2000)Li, Bj{\"o}rkman, Shih, Grossman, Rosenquist,
  Jansson, and Niyogi]{li00a}
Li,~X.-P.; Bj{\"o}rkman,~O.; Shih,~C.; Grossman,~A.~R.; Rosenquist,~M.;
  Jansson,~S.; Niyogi,~K.~K. A Pigment-binding Protein Essential for Regulation
  of Photosynthetic Light Harvesting. \emph{Nature} \textbf{2000}, \emph{403},
  391\relax
\mciteBstWouldAddEndPuncttrue
\mciteSetBstMidEndSepPunct{\mcitedefaultmidpunct}
{\mcitedefaultendpunct}{\mcitedefaultseppunct}\relax
\EndOfBibitem
\bibitem[Correa-Galvis \latin{et~al.}(2016)Correa-Galvis, Poschmann, Melzer,
  St{\"u}hler, and Jahns]{corr16a}
Correa-Galvis,~V.; Poschmann,~G.; Melzer,~M.; St{\"u}hler,~K.; Jahns,~P. PsbS
  Interactions Involved in the Activation of Energy Dissipation in {\emph
  Arabidopsis}. \emph{Nat. Plants} \textbf{2016}, \emph{2}, 1--8\relax
\mciteBstWouldAddEndPuncttrue
\mciteSetBstMidEndSepPunct{\mcitedefaultmidpunct}
{\mcitedefaultendpunct}{\mcitedefaultseppunct}\relax
\EndOfBibitem
\bibitem[Liguori \latin{et~al.}(2019)Liguori, Campos, Baptista, and
  Croce]{ligu19a}
Liguori,~N.; Campos,~S. R.~R.; Baptista,~A.; Croce,~R. Molecular Anatomy of
  Plant Photoprotective Switches: The Sensitivity of Psbs to the Environment,
  Residue by Residue. \emph{J. Phys. Chem. Lett.} \textbf{2019}, \emph{10},
  1737--1742\relax
\mciteBstWouldAddEndPuncttrue
\mciteSetBstMidEndSepPunct{\mcitedefaultmidpunct}
{\mcitedefaultendpunct}{\mcitedefaultseppunct}\relax
\EndOfBibitem
\bibitem[Guardini \latin{et~al.}(2020)Guardini, Bressan, Caferri, Bassi, and
  Dall’Osto]{guar20a}
Guardini,~Z.; Bressan,~M.; Caferri,~R.; Bassi,~R.; Dall’Osto,~L.
  Identification of a Pigment Cluster Catalysing Fast Photoprotective Quenching
  Response in CP29. \emph{Nat. Plants} \textbf{2020}, \emph{6}, 303--313\relax
\mciteBstWouldAddEndPuncttrue
\mciteSetBstMidEndSepPunct{\mcitedefaultmidpunct}
{\mcitedefaultendpunct}{\mcitedefaultseppunct}\relax
\EndOfBibitem
\bibitem[Ruban(2018)]{ruba18a}
Ruban,~A.~V. Light Harvesting Control in Plants. \emph{FEBS Lett.}
  \textbf{2018}, \emph{592}, 3030--3039\relax
\mciteBstWouldAddEndPuncttrue
\mciteSetBstMidEndSepPunct{\mcitedefaultmidpunct}
{\mcitedefaultendpunct}{\mcitedefaultseppunct}\relax
\EndOfBibitem
\bibitem[Dall'Osto \latin{et~al.}(2017)Dall'Osto, Cazzaniga, Bressan,
  Pale{\v{c}}ek, {\v{Z}}idek, Niyogi, Fleming, Zigmantas, and Bassi]{dall17a}
Dall'Osto,~L.; Cazzaniga,~S.; Bressan,~M.; Pale{\v{c}}ek,~D.; {\v{Z}}idek,~K.;
  Niyogi,~K.~K.; Fleming,~G.~R.; Zigmantas,~D.; Bassi,~R. Two Mechanisms for
  Dissipation of Excess Light in Monomeric and Trimeric Light-Harvesting
  Complexes. \emph{Nat. Plants} \textbf{2017}, \emph{3}, 17033\relax
\mciteBstWouldAddEndPuncttrue
\mciteSetBstMidEndSepPunct{\mcitedefaultmidpunct}
{\mcitedefaultendpunct}{\mcitedefaultseppunct}\relax
\EndOfBibitem
\bibitem[Son and Schlau-Cohen(2019)Son, and Schlau-Cohen]{son19a}
Son,~M.; Schlau-Cohen,~G.~S. Flipping a Protein Switch: Carotenoid-Mediated
  Quenching in Plants. \emph{Chem} \textbf{2019}, \emph{5}, 2749--2750\relax
\mciteBstWouldAddEndPuncttrue
\mciteSetBstMidEndSepPunct{\mcitedefaultmidpunct}
{\mcitedefaultendpunct}{\mcitedefaultseppunct}\relax
\EndOfBibitem
\bibitem[{L}iu \latin{et~al.}(2004){L}iu, {Y}an, {W}ang, {K}uang, {Z}hang,
  {G}ui, {A}n, and {C}hang]{liu04a}
{L}iu,~Z.; {Y}an,~H.; {W}ang,~K.; {K}uang,~T.; {Z}hang,~J.; {G}ui,~L.;
  {A}n,~X.; {C}hang,~W. {C}rystal Structure of Spinach Major Light-{H}arvesting
  Complex at 2.72 \r{A} Resolution. \emph{{N}ature} \textbf{2004}, \emph{428},
  287--292\relax
\mciteBstWouldAddEndPuncttrue
\mciteSetBstMidEndSepPunct{\mcitedefaultmidpunct}
{\mcitedefaultendpunct}{\mcitedefaultseppunct}\relax
\EndOfBibitem
\bibitem[M{\"u}h \latin{et~al.}(2010)M{\"u}h, Madjet, and Renger]{muh10a}
M{\"u}h,~F.; Madjet,~M. E.-A.; Renger,~T. Structure-Based Identification of
  Energy Sinks in Plant Light-Harvesting Complex {II}. \emph{J. Phys. Chem. B}
  \textbf{2010}, \emph{114}, 13517--13535\relax
\mciteBstWouldAddEndPuncttrue
\mciteSetBstMidEndSepPunct{\mcitedefaultmidpunct}
{\mcitedefaultendpunct}{\mcitedefaultseppunct}\relax
\EndOfBibitem
\bibitem[M{\"u}h and Renger(2012)M{\"u}h, and Renger]{muh12a}
M{\"u}h,~F.; Renger,~T. Refined Structure-based Simulation of Plant
  Light-harvesting Complex II: Linear Optical Spectra of Trimers and
  Aggregates. \emph{Biochim. Biophys. Acta.-Bioenergetics} \textbf{2012},
  \emph{1817}, 1446--1460\relax
\mciteBstWouldAddEndPuncttrue
\mciteSetBstMidEndSepPunct{\mcitedefaultmidpunct}
{\mcitedefaultendpunct}{\mcitedefaultseppunct}\relax
\EndOfBibitem
\bibitem[Duffy \latin{et~al.}(2013)Duffy, Chmeliov, Macernis, Sulskus,
  Valkunas, and Ruban]{duff13a}
Duffy,~C.; Chmeliov,~J.; Macernis,~M.; Sulskus,~J.; Valkunas,~L.; Ruban,~A.
  Modeling of Fluorescence Quenching by Lutein in the Plant Light-Harvesting
  Complex LHCII. \emph{J. Phys. Chem. B} \textbf{2013}, \emph{117},
  10974--10986\relax
\mciteBstWouldAddEndPuncttrue
\mciteSetBstMidEndSepPunct{\mcitedefaultmidpunct}
{\mcitedefaultendpunct}{\mcitedefaultseppunct}\relax
\EndOfBibitem
\bibitem[Chmeliov \latin{et~al.}(2015)Chmeliov, Bricker, Lo, Jouin, Valkunas,
  Ruban, and Duffy]{chme15a}
Chmeliov,~J.; Bricker,~W.~P.; Lo,~C.; Jouin,~E.; Valkunas,~L.; Ruban,~A.~V.;
  Duffy,~C. D.~P. An {'}All Pigment{'} Model of Excitation Quenching in
  {LHCII}. \emph{Phys. Chem. Chem. Phys.} \textbf{2015}, \emph{17},
  15857--15867\relax
\mciteBstWouldAddEndPuncttrue
\mciteSetBstMidEndSepPunct{\mcitedefaultmidpunct}
{\mcitedefaultendpunct}{\mcitedefaultseppunct}\relax
\EndOfBibitem
\bibitem[Pan \latin{et~al.}(2011)Pan, Li, Wan, Wang, Jia, Hou, Zhao, Zhang, and
  Chang]{pan11a}
Pan,~X.; Li,~M.; Wan,~T.; Wang,~L.; Jia,~C.; Hou,~Z.; Zhao,~X.; Zhang,~J.;
  Chang,~W. Structural Insights into Energy Regulation of Light-Harvesting
  Complex CP29 from Spinach. \emph{Nat. Struct. Mol. Bio.} \textbf{2011},
  \emph{18}, 309--315\relax
\mciteBstWouldAddEndPuncttrue
\mciteSetBstMidEndSepPunct{\mcitedefaultmidpunct}
{\mcitedefaultendpunct}{\mcitedefaultseppunct}\relax
\EndOfBibitem
\bibitem[Wei \latin{et~al.}(2016)Wei, Su, Cao, Liu, Chang, Li, Zhang, and
  Liu]{wei16a}
Wei,~X.; Su,~X.; Cao,~P.; Liu,~X.; Chang,~W.; Li,~M.; Zhang,~X.; Liu,~Z.
  Structure of Spinach Photosystem II-LHCII Supercomplex at 3.2 {\AA}
  Resolution. \emph{Nature} \textbf{2016}, \emph{534}, 69--74\relax
\mciteBstWouldAddEndPuncttrue
\mciteSetBstMidEndSepPunct{\mcitedefaultmidpunct}
{\mcitedefaultendpunct}{\mcitedefaultseppunct}\relax
\EndOfBibitem
\bibitem[M{\"u}h \latin{et~al.}(2014)M{\"u}h, Lindorfer, am~Busch, and
  Renger]{mueh14a}
M{\"u}h,~F.; Lindorfer,~D.; am~Busch,~M.~S.; Renger,~T. Towards a
  Structure-Based Exciton Hamiltonian for the CP29 Antenna of Photosystem II.
  \emph{Phys. Chem. Chem. Phys.} \textbf{2014}, \emph{16}, 11848--11863\relax
\mciteBstWouldAddEndPuncttrue
\mciteSetBstMidEndSepPunct{\mcitedefaultmidpunct}
{\mcitedefaultendpunct}{\mcitedefaultseppunct}\relax
\EndOfBibitem
\bibitem[{J}urinovich \latin{et~al.}(2015){J}urinovich, {V}iani, {P}randi,
  {R}enger, and {M}ennucci]{juri15b}
{J}urinovich,~S.; {V}iani,~L.; {P}randi,~I.~G.; {R}enger,~T.; {M}ennucci,~B.
  {T}owards an Ab Initio Description of the Optical Spectra of
  Light-{H}arvesting Antennae: Application to the {CP}29~{C}omplex of
  Photosystem {II}. \emph{{P}hys. {C}hem. {C}hem. {P}hys.} \textbf{2015},
  \emph{17}, 14405--14416\relax
\mciteBstWouldAddEndPuncttrue
\mciteSetBstMidEndSepPunct{\mcitedefaultmidpunct}
{\mcitedefaultendpunct}{\mcitedefaultseppunct}\relax
\EndOfBibitem
\bibitem[Fox \latin{et~al.}(2018)Fox, {\"U}nl{\"u}, Balevi{\v{c}}ius~Jr,
  Ramdour, Kern, Pan, Li, van Amerongen, and Duffy]{fox18a}
Fox,~K.~F.; {\"U}nl{\"u},~C.; Balevi{\v{c}}ius~Jr,~V.; Ramdour,~B.~N.;
  Kern,~C.; Pan,~X.; Li,~M.; van Amerongen,~H.; Duffy,~C.~D. A Possible
  Molecular Basis for Photoprotection in the Minor Antenna Proteins of Plants.
  \emph{Biochim. Biophys. Acta. -Bioenergetics} \textbf{2018}, \emph{1859},
  471--481\relax
\mciteBstWouldAddEndPuncttrue
\mciteSetBstMidEndSepPunct{\mcitedefaultmidpunct}
{\mcitedefaultendpunct}{\mcitedefaultseppunct}\relax
\EndOfBibitem
\bibitem[Lapillo \latin{et~al.}(2020)Lapillo, Cignoni, Cupellini, and
  Mennucci]{lapi20a}
Lapillo,~M.; Cignoni,~E.; Cupellini,~L.; Mennucci,~B. The Energy Transfer Model
  of Nonphotochemical Quenching: Lessons from the Minor CP29 Antenna Complex of
  Plants. \emph{Biochimica et Biophysica Acta -Bioenergetics} \textbf{2020},
  \emph{1861}, 148282\relax
\mciteBstWouldAddEndPuncttrue
\mciteSetBstMidEndSepPunct{\mcitedefaultmidpunct}
{\mcitedefaultendpunct}{\mcitedefaultseppunct}\relax
\EndOfBibitem
\bibitem[Daskalakis \latin{et~al.}(2019)Daskalakis, Maity, Hart, Stergiannakos,
  Duffy, and Kleinekath{\"{o}}fer]{dask19b}
Daskalakis,~V.; Maity,~S.; Hart,~C.~L.; Stergiannakos,~T.; Duffy,~C. D.~P.;
  Kleinekath{\"{o}}fer,~U. Structural Basis for Allosteric Regulation in the
  Major Antenna Trimer of Photosystem II. \emph{J. Phys. Chem. B}
  \textbf{2019}, \emph{123}, 9609--9615\relax
\mciteBstWouldAddEndPuncttrue
\mciteSetBstMidEndSepPunct{\mcitedefaultmidpunct}
{\mcitedefaultendpunct}{\mcitedefaultseppunct}\relax
\EndOfBibitem
\bibitem[Daskalakis \latin{et~al.}(2019)Daskalakis, Papadatos, and
  Kleinekath{\"{o}}fer]{dask19a}
Daskalakis,~V.; Papadatos,~S.; Kleinekath{\"{o}}fer,~U. Fine Tuning of the
  Photosystem {II} Major Antenna Mobility within the Thylakoid Membrane of
  Higher Plants. \emph{Biochim. Biophys. Acta - Biomembranes} \textbf{2019},
  \emph{1861}, 183059\relax
\mciteBstWouldAddEndPuncttrue
\mciteSetBstMidEndSepPunct{\mcitedefaultmidpunct}
{\mcitedefaultendpunct}{\mcitedefaultseppunct}\relax
\EndOfBibitem
\bibitem[Maity \latin{et~al.}(2019)Maity, Gelessus, Daskalakis, and
  Kleinekath{\"{o}}fer]{mait19a}
Maity,~S.; Gelessus,~A.; Daskalakis,~V.; Kleinekath{\"{o}}fer,~U. On a
  Chlorophyll-Caroteinoid Coupling in LHCII. \emph{Chem. Phys.} \textbf{2019},
  \emph{526}, 110439\relax
\mciteBstWouldAddEndPuncttrue
\mciteSetBstMidEndSepPunct{\mcitedefaultmidpunct}
{\mcitedefaultendpunct}{\mcitedefaultseppunct}\relax
\EndOfBibitem
\bibitem[Daskalakis(2018)]{dask18a}
Daskalakis,~V. Protein-Protein Interactions Within Photosystem {II} under
  Photoprotection: The Synergy Between {CP29} Minor Antenna, Subunit {S}
  {(PsbS)} and Zeaxanthin at All-atom Resolution. \emph{Phys. Chem. Chem.
  Phys.} \textbf{2018}, \emph{20}, 11843--11855\relax
\mciteBstWouldAddEndPuncttrue
\mciteSetBstMidEndSepPunct{\mcitedefaultmidpunct}
{\mcitedefaultendpunct}{\mcitedefaultseppunct}\relax
\EndOfBibitem
\bibitem[Daskalakis \latin{et~al.}(2020)Daskalakis, Papadatos, and
  Stergiannakos]{dask20a}
Daskalakis,~V.; Papadatos,~S.; Stergiannakos,~T. The Conformational Phase Space
  of the Photoprotective Switch in the Major Light Harvesting Complex II.
  \emph{Chem. Comm.} \textbf{2020}, \emph{56}, 11215--11218\relax
\mciteBstWouldAddEndPuncttrue
\mciteSetBstMidEndSepPunct{\mcitedefaultmidpunct}
{\mcitedefaultendpunct}{\mcitedefaultseppunct}\relax
\EndOfBibitem
\bibitem[{K}reisbeck \latin{et~al.}(2014){K}reisbeck, {K}ramer, and
  Aspuru-Guzik]{krei14a}
{K}reisbeck,~C.; {K}ramer,~T.; Aspuru-Guzik,~A. {S}calable {H}igh-{P}erformance
  {A}lgorithm for the {S}imulation of {E}xciton {D}ynamics. {A}pplication to
  the {L}ight-{H}arvesting {C}omplex {II} in the {P}resence of {R}esonant
  {V}ibrational {M}odes. \emph{{J}. {C}hem. {T}heory {C}omput.} \textbf{2014},
  \emph{10}, 4045--4054\relax
\mciteBstWouldAddEndPuncttrue
\mciteSetBstMidEndSepPunct{\mcitedefaultmidpunct}
{\mcitedefaultendpunct}{\mcitedefaultseppunct}\relax
\EndOfBibitem
\bibitem[Roden \latin{et~al.}(2016)Roden, Bennett, and Whaley]{rode16a}
Roden,~J.~J.; Bennett,~D.~I.; Whaley,~K.~B. Long-Range Energy Transport in
  Photosystem II. \emph{J. Chem. Phys.} \textbf{2016}, \emph{144}, 245101\relax
\mciteBstWouldAddEndPuncttrue
\mciteSetBstMidEndSepPunct{\mcitedefaultmidpunct}
{\mcitedefaultendpunct}{\mcitedefaultseppunct}\relax
\EndOfBibitem
\bibitem[{M}ay and {K}\"uhn(2011){M}ay, and {K}\"uhn]{may11}
{M}ay,~V.; {K}\"uhn,~O. \emph{{C}harge and {E}nergy {T}ransfer in {M}olecular
  {S}ystems}, 3rd ed.; Wiley--VCH, 2011\relax
\mciteBstWouldAddEndPuncttrue
\mciteSetBstMidEndSepPunct{\mcitedefaultmidpunct}
{\mcitedefaultendpunct}{\mcitedefaultseppunct}\relax
\EndOfBibitem
\bibitem[{D}amjanovi\'{c} \latin{et~al.}(2002){D}amjanovi\'{c}, {K}osztin,
  {K}leinekath\"ofer, and {S}chulten]{damj02a}
{D}amjanovi\'{c},~A.; {K}osztin,~I.; {K}leinekath\"ofer,~U.; {S}chulten,~K.
  {E}xcitons in a Photosynthetic Light-Harvesting System: {A} Combined
  Molecular Dynamics, Quantum Chemistry and Polaron Model Study. \emph{{P}hys.
  {R}ev. {E}} \textbf{2002}, \emph{65}, 031919\relax
\mciteBstWouldAddEndPuncttrue
\mciteSetBstMidEndSepPunct{\mcitedefaultmidpunct}
{\mcitedefaultendpunct}{\mcitedefaultseppunct}\relax
\EndOfBibitem
\bibitem[{O}lbrich and {K}leinekath\"ofer(2010){O}lbrich, and
  {K}leinekath\"ofer]{olbr10a}
{O}lbrich,~C.; {K}leinekath\"ofer,~U. Time-Dependent Atomistic View on the
  Electronic Relaxation in Light-Harvesting System {II}. \emph{{J}. {P}hys.
  {C}hem. {B}} \textbf{2010}, \emph{114}, 12427--12437\relax
\mciteBstWouldAddEndPuncttrue
\mciteSetBstMidEndSepPunct{\mcitedefaultmidpunct}
{\mcitedefaultendpunct}{\mcitedefaultseppunct}\relax
\EndOfBibitem
\bibitem[{O}lbrich \latin{et~al.}(2011){O}lbrich, {S}tr\"umpfer, {S}chulten,
  and {K}leinekath\"ofer]{olbr11b}
{O}lbrich,~C.; {S}tr\"umpfer,~J.; {S}chulten,~K.; {K}leinekath\"ofer,~U.
  {T}heory and {S}imulation of the {E}nvironmental {E}ffects on {FMO
  E}lectronic {T}ransitions. \emph{{J}. {P}hys. {C}hem. {L}ett.} \textbf{2011},
  \emph{2}, 1771--1776\relax
\mciteBstWouldAddEndPuncttrue
\mciteSetBstMidEndSepPunct{\mcitedefaultmidpunct}
{\mcitedefaultendpunct}{\mcitedefaultseppunct}\relax
\EndOfBibitem
\bibitem[{S}him \latin{et~al.}(2012){S}him, {R}ebentrost, {V}alleau, and
  {A}spuru {G}uzik]{shim11a}
{S}him,~S.; {R}ebentrost,~P.; {V}alleau,~S.; {A}spuru {G}uzik,~A. {A}tomistic
  {S}tudy of the {L}ong-{L}ived {Q}uantum {C}oherences in the
  {F}enna-{M}atthew-{O}lson {C}omplex. \emph{{B}iophys. {J}.} \textbf{2012},
  \emph{102}, 649--660\relax
\mciteBstWouldAddEndPuncttrue
\mciteSetBstMidEndSepPunct{\mcitedefaultmidpunct}
{\mcitedefaultendpunct}{\mcitedefaultseppunct}\relax
\EndOfBibitem
\bibitem[{A}ghtar \latin{et~al.}(2014){A}ghtar, {S}tr\"umpfer, {O}lbrich,
  {S}chulten, and {K}leinekath\"ofer]{aght14a}
{A}ghtar,~M.; {S}tr\"umpfer,~J.; {O}lbrich,~C.; {S}chulten,~K.;
  {K}leinekath\"ofer,~U. {D}ifferent {T}ypes of {V}ibrations {I}nteracting with
  {E}lectronic {E}xcitations in {P}hycoerythrin 545 and
  {F}enna-{M}atthews-{O}lson {A}ntenna {S}ystems. \emph{{J}. {P}hys. {C}hem.
  {L}ett.} \textbf{2014}, \emph{5}, 3131--3137\relax
\mciteBstWouldAddEndPuncttrue
\mciteSetBstMidEndSepPunct{\mcitedefaultmidpunct}
{\mcitedefaultendpunct}{\mcitedefaultseppunct}\relax
\EndOfBibitem
\bibitem[Lee and Coker(2016)Lee, and Coker]{lee16a}
Lee,~M.~K.; Coker,~D.~F. Modeling Electronic-Nuclear Interactions for
  Excitation Energy Transfer Processes in Light-Harvesting Complexes. \emph{J.
  Phys. Chem. Lett.} \textbf{2016}, \emph{7}, 3171--3178\relax
\mciteBstWouldAddEndPuncttrue
\mciteSetBstMidEndSepPunct{\mcitedefaultmidpunct}
{\mcitedefaultendpunct}{\mcitedefaultseppunct}\relax
\EndOfBibitem
\bibitem[Maity \latin{et~al.}(2020)Maity, Bold, Prajapati, Sokolov,
  Kuba{{\v{r}}}, Elstner, and Kleinekath{\"{o}}fer]{mait20a}
Maity,~S.; Bold,~B.~M.; Prajapati,~J.~D.; Sokolov,~M.; Kuba{{\v{r}}},~T.;
  Elstner,~M.; Kleinekath{\"{o}}fer,~U. DFTB/MM Molecular Dynamics Simulations
  of the FMO Light-Harvesting Complex. \emph{J. Phys. Chem. Lett.}
  \textbf{2020}, \emph{11}, 8660--8667\relax
\mciteBstWouldAddEndPuncttrue
\mciteSetBstMidEndSepPunct{\mcitedefaultmidpunct}
{\mcitedefaultendpunct}{\mcitedefaultseppunct}\relax
\EndOfBibitem
\bibitem[{C}urutchet and {M}ennucci(2017){C}urutchet, and {M}ennucci]{curu17a}
{C}urutchet,~C.; {M}ennucci,~B. {Q}uantum Chemical Studies of Light Harvesting.
  \emph{{C}hem. {R}ev.} \textbf{2017}, \emph{117}, 294--343\relax
\mciteBstWouldAddEndPuncttrue
\mciteSetBstMidEndSepPunct{\mcitedefaultmidpunct}
{\mcitedefaultendpunct}{\mcitedefaultseppunct}\relax
\EndOfBibitem
\bibitem[Lee \latin{et~al.}(2016)Lee, Huo, and Coker]{lee16c}
Lee,~M.~K.; Huo,~P.; Coker,~D.~F. Semiclassical Path Integral Dynamics:
  Photosynthetic Energy Transfer with Realistic Environment Interactions.
  \emph{Annu. Rev. Phys. Chem.} \textbf{2016}, \emph{67}, 639--668\relax
\mciteBstWouldAddEndPuncttrue
\mciteSetBstMidEndSepPunct{\mcitedefaultmidpunct}
{\mcitedefaultendpunct}{\mcitedefaultseppunct}\relax
\EndOfBibitem
\bibitem[Kim and Rhee(2016)Kim, and Rhee]{kim16a}
Kim,~C.~W.; Rhee,~Y.~M. Constructing an Interpolated Potential Energy Surface
  of a Large Molecule: A Case Study with Bacteriochlorophyll a Model in the
  {F}enna-{M}atthews-{O}lson Complex. \emph{J. Chem. Theory Comput.}
  \textbf{2016}, \emph{12}, 5235--5246\relax
\mciteBstWouldAddEndPuncttrue
\mciteSetBstMidEndSepPunct{\mcitedefaultmidpunct}
{\mcitedefaultendpunct}{\mcitedefaultseppunct}\relax
\EndOfBibitem
\bibitem[Kim \latin{et~al.}(2018)Kim, Choi, and Rhee]{kim18a}
Kim,~C.~W.; Choi,~B.; Rhee,~Y.~M. Excited State Energy Fluctuations in the
  Fenna-Matthews-Olson Complex from Molecular Dynamics Simulations with
  Interpolated Chromophore Potentials. \emph{Phys. Chem. Chem. Phys.}
  \textbf{2018}, \emph{20}, 3310\relax
\mciteBstWouldAddEndPuncttrue
\mciteSetBstMidEndSepPunct{\mcitedefaultmidpunct}
{\mcitedefaultendpunct}{\mcitedefaultseppunct}\relax
\EndOfBibitem
\bibitem[Padula \latin{et~al.}(2017)Padula, Lee, Claridge, and Troisi]{padu17a}
Padula,~D.; Lee,~M.~H.; Claridge,~K.; Troisi,~A. Chromophore-Dependent
  Intramolecular Exciton-Vibrational Coupling in the FMO Complex:
  Quantification and Importance for Exciton Dynamics. \emph{J. Phys. Chem. B}
  \textbf{2017}, \emph{121}, 10026--10035\relax
\mciteBstWouldAddEndPuncttrue
\mciteSetBstMidEndSepPunct{\mcitedefaultmidpunct}
{\mcitedefaultendpunct}{\mcitedefaultseppunct}\relax
\EndOfBibitem
\bibitem[{R}osnik and {C}urutchet(2015){R}osnik, and {C}urutchet]{rosn15a}
{R}osnik,~A.~M.; {C}urutchet,~C. {T}heoretical {C}haracterization of the
  {S}pectral {D}ensity of the {W}ater-{S}oluble {C}hlorophyll-{B}inding
  {P}rotein from {C}ombined {Q}uantum {M}echanics/{M}olecular {M}echanics
  {M}olecular {D}ynamics {S}imulations. \emph{{J}. {C}hem. {T}heory {C}omput.}
  \textbf{2015}, \emph{11}, 5826--5837\relax
\mciteBstWouldAddEndPuncttrue
\mciteSetBstMidEndSepPunct{\mcitedefaultmidpunct}
{\mcitedefaultendpunct}{\mcitedefaultseppunct}\relax
\EndOfBibitem
\bibitem[Blau \latin{et~al.}(2018)Blau, Bennett, Kreisbeck, Scholes, and
  Aspuru-Guzik]{blau18a}
Blau,~S.~M.; Bennett,~D. I.~G.; Kreisbeck,~C.; Scholes,~G.~D.; Aspuru-Guzik,~A.
  Local Protein Solvation Drives Direct Down-Conversion in Phycobiliprotein
  PC645 Via Incoherent Vibronic Transport. \emph{{P}roc. {N}atl. {A}cad. {S}ci.
  {USA}} \textbf{2018}, \emph{115}, E3342--E3350\relax
\mciteBstWouldAddEndPuncttrue
\mciteSetBstMidEndSepPunct{\mcitedefaultmidpunct}
{\mcitedefaultendpunct}{\mcitedefaultseppunct}\relax
\EndOfBibitem
\bibitem[Elstner \latin{et~al.}(1998)Elstner, Porezag, Jungnickel, Elsner,
  Haugk, Frauenheim, Suhai, and Seifert]{elst98a}
Elstner,~M.; Porezag,~D.; Jungnickel,~G.; Elsner,~J.; Haugk,~M.;
  Frauenheim,~T.; Suhai,~S.; Seifert,~G. Self-consistent-charge
  Density-functional Tight-binding Method for Simulations of Complex Materials
  Properties. \emph{Phys. Rev. B} \textbf{1998}, \emph{58}, 7260--7268\relax
\mciteBstWouldAddEndPuncttrue
\mciteSetBstMidEndSepPunct{\mcitedefaultmidpunct}
{\mcitedefaultendpunct}{\mcitedefaultseppunct}\relax
\EndOfBibitem
\bibitem[Maity \latin{et~al.}(2021)Maity, Daskalakis, Elstner, and
  Kleinekath{\"{o}}fer]{mait21a}
Maity,~S.; Daskalakis,~V.; Elstner,~M.; Kleinekath{\"{o}}fer,~U. Multiscale
  QM/MM Molecular Dynamics Simulations of the Trimeric Major Light-Harvesting
  Complex II. \emph{Phys. Chem. Chem. Phys.} \textbf{2021}, \emph{23},
  7407--7417\relax
\mciteBstWouldAddEndPuncttrue
\mciteSetBstMidEndSepPunct{\mcitedefaultmidpunct}
{\mcitedefaultendpunct}{\mcitedefaultseppunct}\relax
\EndOfBibitem
\bibitem[Duan \latin{et~al.}(2003)Duan, Wu, Chowdhury, Lee, Xiong, Zhang, Yang,
  Cieplak, Luo, Lee, \latin{et~al.} others]{duan03a}
Duan,~Y.; Wu,~C.; Chowdhury,~S.; Lee,~M.~C.; Xiong,~G.; Zhang,~W.; Yang,~R.;
  Cieplak,~P.; Luo,~R.; Lee,~T., \latin{et~al.}  A Point-Charge Force Field for
  Molecular Mechanics Simulations of Proteins Based on Condensed-Phase Quantum
  Mechanical Calculations. \emph{J. Comput. Chem.} \textbf{2003}, \emph{24},
  1999--2012\relax
\mciteBstWouldAddEndPuncttrue
\mciteSetBstMidEndSepPunct{\mcitedefaultmidpunct}
{\mcitedefaultendpunct}{\mcitedefaultseppunct}\relax
\EndOfBibitem
\bibitem[Abraham \latin{et~al.}(2015)Abraham, Murtola, Schulz, P{\'a}ll, Smith,
  Hess, and Lindahl]{abra15a}
Abraham,~M.~J.; Murtola,~T.; Schulz,~R.; P{\'a}ll,~S.; Smith,~J.~C.; Hess,~B.;
  Lindahl,~E. {GROMACS}: High Performance Molecular Simulations through
  Multi-Level Parallelism from Laptops to Supercomputers. \emph{SoftwareX}
  \textbf{2015}, \emph{1}, 19--25\relax
\mciteBstWouldAddEndPuncttrue
\mciteSetBstMidEndSepPunct{\mcitedefaultmidpunct}
{\mcitedefaultendpunct}{\mcitedefaultseppunct}\relax
\EndOfBibitem
\bibitem[{C}eccarelli \latin{et~al.}(2003){C}eccarelli, {P}rocacci, and
  {M}archi]{cecc03a}
{C}eccarelli,~M.; {P}rocacci,~P.; {M}archi,~M. {A}n {A}b {I}nitio {F}orce
  {F}ield for the {C}ofactors of {B}acterial {P}hotosynthesis. \emph{{J}.
  {C}omput. {C}hem.} \textbf{2003}, \emph{24}, 129--132\relax
\mciteBstWouldAddEndPuncttrue
\mciteSetBstMidEndSepPunct{\mcitedefaultmidpunct}
{\mcitedefaultendpunct}{\mcitedefaultseppunct}\relax
\EndOfBibitem
\bibitem[{Z}hang \latin{et~al.}(2012){Z}hang, {S}ilva, {Y}an, and
  {H}uang]{zhan12a}
{Z}hang,~L.; {S}ilva,~D.-A.; {Y}an,~Y.; {H}uang,~X. {F}orce Field Development
  for Cofactors in the Photosystem {II}. \emph{{J}. {C}omput. {C}hem.}
  \textbf{2012}, \emph{33}, 1969--1980\relax
\mciteBstWouldAddEndPuncttrue
\mciteSetBstMidEndSepPunct{\mcitedefaultmidpunct}
{\mcitedefaultendpunct}{\mcitedefaultseppunct}\relax
\EndOfBibitem
\bibitem[{P}randi \latin{et~al.}(2016){P}randi, {V}iani, {A}ndreussi, and
  {M}ennucci]{pran16a}
{P}randi,~I.~G.; {V}iani,~L.; {A}ndreussi,~O.; {M}ennucci,~B. {C}ombining
  Classical Molecular Dynamics and Quantum Mechanical Methods for the
  Description of Electronic Excitations: {T}he Case of Carotenoids. \emph{{J}.
  {C}omput. {C}hem.} \textbf{2016}, \emph{37}, 981--991\relax
\mciteBstWouldAddEndPuncttrue
\mciteSetBstMidEndSepPunct{\mcitedefaultmidpunct}
{\mcitedefaultendpunct}{\mcitedefaultseppunct}\relax
\EndOfBibitem
\bibitem[{J}o \latin{et~al.}(2008){J}o, {K}im, {I}yer, and {I}m]{jo08a}
{J}o,~S.; {K}im,~T.; {I}yer,~V.~G.; {I}m,~W. {CHARMM-GUI}: A Web-Based
  Graphical User Interface for {CHARMM}. \emph{{J}. {C}omput. {C}hem.}
  \textbf{2008}, \emph{29}, 1859--1865\relax
\mciteBstWouldAddEndPuncttrue
\mciteSetBstMidEndSepPunct{\mcitedefaultmidpunct}
{\mcitedefaultendpunct}{\mcitedefaultseppunct}\relax
\EndOfBibitem
\bibitem[Da~Silva and Vranken(2012)Da~Silva, and Vranken]{dasi12a}
Da~Silva,~A. W.~S.; Vranken,~W.~F. ACPYPE-Antechamber Python Parser Interface.
  \emph{BMC Res. Notes} \textbf{2012}, \emph{5}, 367\relax
\mciteBstWouldAddEndPuncttrue
\mciteSetBstMidEndSepPunct{\mcitedefaultmidpunct}
{\mcitedefaultendpunct}{\mcitedefaultseppunct}\relax
\EndOfBibitem
\bibitem[{M}adjet \latin{et~al.}(2006){M}adjet, {A}bdurahman, and
  {R}enger]{madj06a}
{M}adjet,~M.~E.; {A}bdurahman,~A.; {R}enger,~T. {I}ntermolecular Coulomb
  Couplings from Ab Initio Electrostatic Potentials: {A}pplication to Optical
  Transitions of Strongly Coupled Pigments in Photosynthetic Antennae and
  Reaction Centers. \emph{{J}. {P}hys. {C}hem. {B}} \textbf{2006}, \emph{110},
  17268--81\relax
\mciteBstWouldAddEndPuncttrue
\mciteSetBstMidEndSepPunct{\mcitedefaultmidpunct}
{\mcitedefaultendpunct}{\mcitedefaultseppunct}\relax
\EndOfBibitem
\bibitem[{O}lbrich \latin{et~al.}(2011){O}lbrich, {J}ansen, {L}iebers,
  {A}ghtar, {S}tr\"umpfer, {S}chulten, {K}noester, and
  {K}leinekath\"ofer]{olbr11a}
{O}lbrich,~C.; {J}ansen,~T. L.~C.; {L}iebers,~J.; {A}ghtar,~M.;
  {S}tr\"umpfer,~J.; {S}chulten,~K.; {K}noester,~J.; {K}leinekath\"ofer,~U.
  {F}rom {A}tomistic {M}odeling to {E}xcitation {D}ynamics and
  {T}wo-{D}imensional {S}pectra of the {FMO L}ight-{H}arvesting {C}omplex.
  \emph{{J}. {P}hys. {C}hem. {B}} \textbf{2011}, \emph{115}, 8609--8621\relax
\mciteBstWouldAddEndPuncttrue
\mciteSetBstMidEndSepPunct{\mcitedefaultmidpunct}
{\mcitedefaultendpunct}{\mcitedefaultseppunct}\relax
\EndOfBibitem
\bibitem[Su \latin{et~al.}(2017)Su, Ma, Wei, Cao, Zhu, Chang, Liu, Zhang, and
  Li]{su17a}
Su,~X.; Ma,~J.; Wei,~X.; Cao,~P.; Zhu,~D.; Chang,~W.; Liu,~Z.; Zhang,~X.;
  Li,~M. Structure and Assembly Mechanism of Plant C2S2M2-type PSII-LHCII
  Supercomplex. \emph{Science} \textbf{2017}, \emph{357}, 815--820\relax
\mciteBstWouldAddEndPuncttrue
\mciteSetBstMidEndSepPunct{\mcitedefaultmidpunct}
{\mcitedefaultendpunct}{\mcitedefaultseppunct}\relax
\EndOfBibitem
\bibitem[Bold \latin{et~al.}(2020)Bold, Sokolov, Maity, Wanko, Dohmen, Kranz,
  Kleinekath{\"o}fer, H{\"{o}}fener, and Elstner]{bold20a}
Bold,~B.~M.; Sokolov,~M.; Maity,~S.; Wanko,~M.; Dohmen,~P.~M.; Kranz,~J.~J.;
  Kleinekath{\"o}fer,~U.; H{\"{o}}fener,~S.; Elstner,~M. Benchmark and
  Performance of Long-Range Corrected Time-Dependent Density Functional Tight
  Binding (LC-TD-DFTB) on Rhodopsins and Light-Harvesting Complexes.
  \emph{Phys. Chem. Chem. Phys.} \textbf{2020}, \emph{22}, 10500--10518\relax
\mciteBstWouldAddEndPuncttrue
\mciteSetBstMidEndSepPunct{\mcitedefaultmidpunct}
{\mcitedefaultendpunct}{\mcitedefaultseppunct}\relax
\EndOfBibitem
\bibitem[{G}aus \latin{et~al.}(2011){G}aus, {C}ui, and {E}lstner]{gaus11a}
{G}aus,~M.; {C}ui,~Q.; {E}lstner,~M. {DFTB}3: {E}xtension of the
  {S}elf-{C}onsistent-{C}harge {D}ensity-{F}unctional {T}ight-{B}inding
  {M}ethod ({SCC-{DFT}B}). \emph{{J}. {C}hem. {T}heory {C}omput.}
  \textbf{2011}, \emph{7}, 931--948\relax
\mciteBstWouldAddEndPuncttrue
\mciteSetBstMidEndSepPunct{\mcitedefaultmidpunct}
{\mcitedefaultendpunct}{\mcitedefaultseppunct}\relax
\EndOfBibitem
\bibitem[Gaus \latin{et~al.}(2013)Gaus, Goez, and Elstner]{gaus13a}
Gaus,~M.; Goez,~A.; Elstner,~M. Parametrization and Benchmark of {DFTB3} for
  Organic Molecules. \emph{J. Chem. Theory Comput.} \textbf{2013}, \emph{9},
  338--354\relax
\mciteBstWouldAddEndPuncttrue
\mciteSetBstMidEndSepPunct{\mcitedefaultmidpunct}
{\mcitedefaultendpunct}{\mcitedefaultseppunct}\relax
\EndOfBibitem
\bibitem[Kuba{\v{r}} \latin{et~al.}(2015)Kuba{\v{r}}, Welke, and
  Groenhof]{kuba15b}
Kuba{\v{r}},~T.; Welke,~K.; Groenhof,~G. New QM/MM Implementation of the DFTB3
  Method in the Gromacs Package. \emph{J. Comput. Chem.} \textbf{2015},
  \emph{36}, 1978--1989\relax
\mciteBstWouldAddEndPuncttrue
\mciteSetBstMidEndSepPunct{\mcitedefaultmidpunct}
{\mcitedefaultendpunct}{\mcitedefaultseppunct}\relax
\EndOfBibitem
\bibitem[Hourahine \latin{et~al.}(2020)Hourahine, Aradi, Blum, Bonaf{\'{e}},
  Buccheri, Camacho, Cevallos, Deshaye, Dumitric{\u{a}}, Dominguez, Ehlert,
  Elstner, Van Der~Heide, Hermann, Irle, Kranz, K{\"{o}}hler, Kowalczyk,
  Kuba{{\v{r}}}, Lee, Lutsker, Maurer, Min, Mitchell, Negre, Niehaus,
  Niklasson, Page, Pecchia, Penazzi, Persson, {{\v{R}}}ez{\'{a}}{\v{c}},
  S{\'{a}}nchez, Sternberg, St{\"{o}}hr, Stuckenberg, Tkatchenko, Yu, and
  Frauenheim]{hour20a}
Hourahine,~B. \latin{et~al.}  DFTB+, a Software Package for Efficient
  Approximate Density Functional Theory Based Atomistic Simulations. \emph{J.
  Chem. Phys.} \textbf{2020}, \emph{152}, 124101\relax
\mciteBstWouldAddEndPuncttrue
\mciteSetBstMidEndSepPunct{\mcitedefaultmidpunct}
{\mcitedefaultendpunct}{\mcitedefaultseppunct}\relax
\EndOfBibitem
\bibitem[Kranz \latin{et~al.}(2017)Kranz, Elstner, Aradi, Frauenheim, Lutsker,
  Garcia, and Niehaus]{kran17a}
Kranz,~J.~J.; Elstner,~M.; Aradi,~B.; Frauenheim,~T.; Lutsker,~V.;
  Garcia,~A.~D.; Niehaus,~T.~A. Time-Dependent Extension of the Long-Range
  Corrected Density Functional Based Tight-Binding Method. \emph{J. Chem.
  Theory Comput.} \textbf{2017}, \emph{13}, 1737--1747\relax
\mciteBstWouldAddEndPuncttrue
\mciteSetBstMidEndSepPunct{\mcitedefaultmidpunct}
{\mcitedefaultendpunct}{\mcitedefaultseppunct}\relax
\EndOfBibitem
\bibitem[Lu and Chen(2012)Lu, and Chen]{lu12a}
Lu,~T.; Chen,~F. Multiwfn: A Multifunctional Wavefunction Analyzer. \emph{J.
  Comput. Chem.} \textbf{2012}, \emph{33}, 580--592\relax
\mciteBstWouldAddEndPuncttrue
\mciteSetBstMidEndSepPunct{\mcitedefaultmidpunct}
{\mcitedefaultendpunct}{\mcitedefaultseppunct}\relax
\EndOfBibitem
\bibitem[Neese(2018)]{nees18a}
Neese,~F. Software Update: The {ORCA} Program System, Version 4.0. \emph{WIREs
  Comput. Mol. Sci.} \textbf{2018}, \emph{8}, e1327\relax
\mciteBstWouldAddEndPuncttrue
\mciteSetBstMidEndSepPunct{\mcitedefaultmidpunct}
{\mcitedefaultendpunct}{\mcitedefaultseppunct}\relax
\EndOfBibitem
\bibitem[Knox and Spring(2003)Knox, and Spring]{knox03a}
Knox,~R.~S.; Spring,~B.~Q. Dipole Strengths in the Chlorophylls.
  \emph{Photochem. Photobiol.} \textbf{2003}, \emph{77}, 497--501\relax
\mciteBstWouldAddEndPuncttrue
\mciteSetBstMidEndSepPunct{\mcitedefaultmidpunct}
{\mcitedefaultendpunct}{\mcitedefaultseppunct}\relax
\EndOfBibitem
\bibitem[Aghtar \latin{et~al.}(2012)Aghtar, Liebers, Str{\"{u}}mpfer, Schulten,
  and Kleinekath{\"{o}}fer]{aght12a}
Aghtar,~M.; Liebers,~J.; Str{\"{u}}mpfer,~J.; Schulten,~K.;
  Kleinekath{\"{o}}fer,~U. {J}uxtaposing {D}ensity {M}atrix and {C}lassical
  {P}ath-Based {W}ave {P}acket {D}ynamics. \emph{{J}. {C}hem. {P}hys.}
  \textbf{2012}, \emph{136}, 214101\relax
\mciteBstWouldAddEndPuncttrue
\mciteSetBstMidEndSepPunct{\mcitedefaultmidpunct}
{\mcitedefaultendpunct}{\mcitedefaultseppunct}\relax
\EndOfBibitem
\bibitem[Jansen(2018)]{jans18a}
Jansen,~T. L.~C. Simple Quantum Dynamics with Thermalization. \emph{J. Phys.
  Chem. A} \textbf{2018}, \emph{122}, 172--183\relax
\mciteBstWouldAddEndPuncttrue
\mciteSetBstMidEndSepPunct{\mcitedefaultmidpunct}
{\mcitedefaultendpunct}{\mcitedefaultseppunct}\relax
\EndOfBibitem
\bibitem[Aghtar \latin{et~al.}(2017)Aghtar, Kleinekath{\"{o}}fer, Curutchet,
  and Mennucci]{aght17a}
Aghtar,~M.; Kleinekath{\"{o}}fer,~U.; Curutchet,~C.; Mennucci,~B. Impact Of
  Electronic Fluctuations And Their Description On The Exciton Dynamics In The
  Light-Harvesting Complex {PE545}. \emph{J. Phys. Chem. B} \textbf{2017},
  \emph{121}, 1330--1339\relax
\mciteBstWouldAddEndPuncttrue
\mciteSetBstMidEndSepPunct{\mcitedefaultmidpunct}
{\mcitedefaultendpunct}{\mcitedefaultseppunct}\relax
\EndOfBibitem
\bibitem[{V}alleau \latin{et~al.}(2012){V}alleau, {E}isfeld, and {A}spuru
  {G}uzik]{vall12a}
{V}alleau,~S.; {E}isfeld,~A.; {A}spuru {G}uzik,~A. {O}n the {A}lternatives for
  {B}ath {C}orrelators and {S}pectral {D}ensities from {M}ixed
  {Q}uantum-{C}lassical {S}imulations. \emph{{J}. {C}hem. {P}hys.}
  \textbf{2012}, \emph{137}, 224103--13\relax
\mciteBstWouldAddEndPuncttrue
\mciteSetBstMidEndSepPunct{\mcitedefaultmidpunct}
{\mcitedefaultendpunct}{\mcitedefaultseppunct}\relax
\EndOfBibitem
\bibitem[{S}choles \latin{et~al.}(2007){S}choles, {C}urutchet, {M}ennucci,
  {C}ammi, and {J}.{T}omasi]{scho07a}
{S}choles,~G.~D.; {C}urutchet,~C.; {M}ennucci,~B.; {C}ammi,~R.; {J}.{T}omasi,
  How Solvent Controls Electronic Energy Transfer and Light Harvesting.
  \emph{{J}. {P}hys. {C}hem. {B}} \textbf{2007}, \emph{111}, 13253--13265\relax
\mciteBstWouldAddEndPuncttrue
\mciteSetBstMidEndSepPunct{\mcitedefaultmidpunct}
{\mcitedefaultendpunct}{\mcitedefaultseppunct}\relax
\EndOfBibitem
\bibitem[Jassas \latin{et~al.}(2018)Jassas, Chen, Khmelnitskiy, Casazza,
  Santabarbara, and Jankowiak]{jass18a}
Jassas,~M.; Chen,~J.; Khmelnitskiy,~A.; Casazza,~A.~P.; Santabarbara,~S.;
  Jankowiak,~R. Structure-Based Exciton Hamiltonian and Dynamics for the
  Reconstituted Wild-Type CP29 Protein Antenna Complex of the Photosystem II.
  \emph{J. Phys. Chem. B} \textbf{2018}, \emph{122}, 4611--4624\relax
\mciteBstWouldAddEndPuncttrue
\mciteSetBstMidEndSepPunct{\mcitedefaultmidpunct}
{\mcitedefaultendpunct}{\mcitedefaultseppunct}\relax
\EndOfBibitem
\bibitem[Mascoli \latin{et~al.}(2020)Mascoli, Novoderezhkin, Liguori, Xu, and
  Croce]{masc20a}
Mascoli,~V.; Novoderezhkin,~V.; Liguori,~N.; Xu,~P.; Croce,~R. Design
  Principles of Solar Light Harvesting in Plants: Functional Architecture of
  the Monomeric Antenna CP29. \emph{Biochim. Biophys. Acta - Bioenergetics}
  \textbf{2020}, \emph{1861}, 148156\relax
\mciteBstWouldAddEndPuncttrue
\mciteSetBstMidEndSepPunct{\mcitedefaultmidpunct}
{\mcitedefaultendpunct}{\mcitedefaultseppunct}\relax
\EndOfBibitem
\bibitem[Bennett \latin{et~al.}(2013)Bennett, Amarnath, and Fleming]{benn13a}
Bennett,~D.~I.; Amarnath,~K.; Fleming,~G.~R. A Structure-Based Model of Energy
  Transfer Reveals the Principles of Light Harvesting in Photosystem II
  Supercomplexes. \emph{J. Am. Chem. Soc.} \textbf{2013}, \emph{135},
  9164--9173\relax
\mciteBstWouldAddEndPuncttrue
\mciteSetBstMidEndSepPunct{\mcitedefaultmidpunct}
{\mcitedefaultendpunct}{\mcitedefaultseppunct}\relax
\EndOfBibitem
\bibitem[{N}ovoderezhkin \latin{et~al.}(2004){N}ovoderezhkin, {P}alacios, van
  {A}merongen, and van {G}rondelle]{novo04a}
{N}ovoderezhkin,~V.; {P}alacios,~M.~A.; van {A}merongen,~H.; van
  {G}rondelle,~R. {E}nergy-{T}ransfer {D}ynamics in the {LHCII C}omplex of
  {H}igher {P}lants: {M}odified {R}edfield {A}pproach. \emph{{J}. {P}hys.
  {C}hem. {B}} \textbf{2004}, \emph{108}, 10363\relax
\mciteBstWouldAddEndPuncttrue
\mciteSetBstMidEndSepPunct{\mcitedefaultmidpunct}
{\mcitedefaultendpunct}{\mcitedefaultseppunct}\relax
\EndOfBibitem
\bibitem[{K}ell \latin{et~al.}(2013){K}ell, {F}eng, {R}eppert, and
  {J}ankowiak]{kell13a}
{K}ell,~A.; {F}eng,~X.; {R}eppert,~M.; {J}ankowiak,~R. {O}n the {S}hape of the
  {P}honon {S}pectral {D}ensity in {P}hotosynthetic {C}omplexes. \emph{{J}.
  {P}hys. {C}hem. {B}} \textbf{2013}, \emph{117}, 7317--7323\relax
\mciteBstWouldAddEndPuncttrue
\mciteSetBstMidEndSepPunct{\mcitedefaultmidpunct}
{\mcitedefaultendpunct}{\mcitedefaultseppunct}\relax
\EndOfBibitem
\bibitem[{C}handrasekaran \latin{et~al.}(2015){C}handrasekaran, {A}ghtar,
  {V}alleau, Aspuru-Guzik, and {K}leinekath\"ofer]{chan15a}
{C}handrasekaran,~S.; {A}ghtar,~M.; {V}alleau,~S.; Aspuru-Guzik,~A.;
  {K}leinekath\"ofer,~U. {I}nfluence of {F}orce {F}ields and {Q}uantum
  {C}hemistry {A}pproach on {S}pectral {D}ensities of {BC}hl a in {S}olution
  and in {FMO P}roteins. \emph{{J}. {P}hys. {C}hem. {B}} \textbf{2015},
  \emph{119}, 9995--10004\relax
\mciteBstWouldAddEndPuncttrue
\mciteSetBstMidEndSepPunct{\mcitedefaultmidpunct}
{\mcitedefaultendpunct}{\mcitedefaultseppunct}\relax
\EndOfBibitem
\bibitem[Jorgensen and Tirado-Rives(1988)Jorgensen, and Tirado-Rives]{jorg88a}
Jorgensen,~W.~L.; Tirado-Rives,~J. The {OPLS} [Optimized Potentials for Liquid
  Simulations] Potential Functions for Proteins, Energy Minimizations for
  Crystals of Cyclic Peptides and Crambin. \emph{J. Am. Chem. Soc.}
  \textbf{1988}, \emph{110}, 1657--1666\relax
\mciteBstWouldAddEndPuncttrue
\mciteSetBstMidEndSepPunct{\mcitedefaultmidpunct}
{\mcitedefaultendpunct}{\mcitedefaultseppunct}\relax
\EndOfBibitem
\bibitem[{R}\"atsep and {F}reiberg(2007){R}\"atsep, and {F}reiberg]{raet07a}
{R}\"atsep,~M.; {F}reiberg,~A. {E}lectron-Phonon and Vibronic Couplings in the
  {FMO} Bacteriochlorophyll a Antenna Complex Studied by Difference
  Fluorescence Line Narrowing. \emph{J. Lumin.} \textbf{2007}, \emph{127},
  251--259\relax
\mciteBstWouldAddEndPuncttrue
\mciteSetBstMidEndSepPunct{\mcitedefaultmidpunct}
{\mcitedefaultendpunct}{\mcitedefaultseppunct}\relax
\EndOfBibitem
\bibitem[Singharoy \latin{et~al.}(2019)Singharoy, Maffeo, Delgado-Magnero,
  Swainsbury, {\c S}ener, Kleinekath{\"{o}}fer, Vant, Nguyen, Hitchcock,
  Isralewitz, Teo, Chandler, Stone, Phillips, Pogorelov, Mallus, Chipot,
  Luthey-Schulten, Tieleman, Hunter, Tajkhorshid, Aksimentiev, and
  Schulten]{sing19a}
Singharoy,~A. \latin{et~al.}  Atoms to Phenotypes: Molecular Design Principles
  of Cellular Energy Metabolism. \emph{Cell} \textbf{2019}, \emph{179},
  1098--1111.e23\relax
\mciteBstWouldAddEndPuncttrue
\mciteSetBstMidEndSepPunct{\mcitedefaultmidpunct}
{\mcitedefaultendpunct}{\mcitedefaultseppunct}\relax
\EndOfBibitem
\end{mcitethebibliography}
\providecommand{\latin}[1]{#1}
\makeatletter
\providecommand{\doi}
  {\begingroup\let\do\@makeother\dospecials
  \catcode`\{=1 \catcode`\}=2 \doi@aux}
\providecommand{\doi@aux}[1]{\endgroup\texttt{#1}}
\makeatother
\providecommand*\mcitethebibliography{\thebibliography}
\csname @ifundefined\endcsname{endmcitethebibliography}
  {\let\endmcitethebibliography\endthebibliography}{}

\end{document}